\documentclass[floatfix,amsmath,amssymb,superscriptaddress,10pt,twocolumn,showkeys]{revtex4-1}
\usepackage{hyperref}
\usepackage{booktabs,makecell}
\usepackage[flushleft]{threeparttable}

\usepackage{textgreek}

\usepackage{lipsum}  

\usepackage{mathtools}
\usepackage{color,soul}

\usepackage{miller}

\usepackage{natbib}
\DeclareUnicodeCharacter{0394}{$\Delta$}

\newcommand\pymulskips{\textsc{PyMulSKIPS}~}
\newcommand\mulskips{\textsc{MulSKIPS}~}
\newcommand\fenics{\textsc{FEniCS}~}
\newcommand\dolfin{\textsc{Dolfin}~}

\newcommand\sige[2]{\textrm{Si}\textsubscript{#1}\textrm{Ge}\textsubscript{#2}}

\newcommand\Eref[1]{Equation~\eqref{#1}}
\newcommand\Fref[1]{Figure~\ref{#1}}

\newcommand\Tref[1]{Table~\ref{#1}}

\begin{document}

\title{Atomistic insights into ultrafast SiGe nanoprocessing}

\author{Gaetano Calogero}
\email{E-mail: gaetano.calogero@imm.cnr.it (corresponding author)}
\affiliation{CNR IMM, Z.I. VIII Strada 5, 95121 Catania, Italy}

\author{Domenica Raciti}
\affiliation{STMicroelectronics, Stradale Primosole 50, 95121 Catania, Italy}

\author{Damiano Ricciarelli}
\affiliation{CNR IMM, Z.I. VIII Strada 5, 95121 Catania, Italy}

\author{Pablo Acosta-Alba}
\affiliation{Université Grenoble Alpes, CEA, LETI, 38000 Grenoble, France}

\author{Fuccio Cristiano}
\affiliation{LAAS-CNRS, Université de Toulouse, 31400 Toulouse, France}

\author{Richard Daubriac} 
\affiliation{LAAS-CNRS, Université de Toulouse, 31400 Toulouse, France}

\author{Remi Demoulin} 
\affiliation{Université Rouen Normandie, Saint Etienne Du Rouvray, France}

\author{Ioannis Deretzis}
\affiliation{CNR IMM, Z.I. VIII Strada 5, 95121 Catania, Italy}

\author{Giuseppe Fisicaro}
\affiliation{CNR IMM, Z.I. VIII Strada 5, 95121 Catania, Italy}

\author{Jean-Michel Hartmann}
\affiliation{Université Grenoble Alpes, CEA, LETI, 38000 Grenoble, France}

\author{S{\'{e}}bastien Kerdil{\`{e}}s}
\affiliation{Université Grenoble Alpes, CEA, LETI, 38000 Grenoble, France}

\author{Antonino La Magna}
\email{E-mail: antonino.lamagna@imm.cnr.it}
\affiliation{CNR IMM, Z.I. VIII Strada 5, 95121 Catania, Italy}

\begin{abstract}
Controlling ultrafast material transformations with atomic precision is essential for future nanotechnology. Pulsed laser annealing (LA), inducing extremely rapid and localized phase transitions, is a powerful way to achieve this, but it requires careful optimization together with the appropriate system design. We present a multiscale LA computational framework able to simulate atom-by-atom the highly out-of-equilibrium kinetics of a material as it interacts with the laser, including effects of structural disorder. By seamlessly coupling a macroscale continuum solver to a nanoscale super-lattice Kinetic Monte Carlo code, this method overcomes the limits of state-of-the-art continuum-based tools. We exploit it to investigate nontrivial changes in composition, morphology and quality of laser-annealed SiGe alloys. Validations against experiments and phase-field simulations, as well as advanced applications to strained, defected, nanostructured and confined SiGe are presented, highlighting the importance of a multiscale atomistic-continuum approach. Current applicability and potential generalization routes are finally discussed.
\end{abstract}


\maketitle

As material science roadmaps relentlessly pursue the digital, sustainability and quantum paradigms, understanding and harnessing ultrafast transformations at the atomic scale is becoming increasingly crucial for the atom-by-atom control of nanosystems and their integration as building blocks into meso- and macroscale systems \cite{MaterialsRoadmap2030, Giustino2020, IRDS2022, Alcorn2023, CristianoLaMagna2021}. Laser annealing (LA) using excimer pulses is an excellent and long-standing way of inducing and investigating such transformations, as it enables localized energy absorption, heating and melting over nm-sized subportions of the material in extremely short time (from tens to hundrends of ns) \cite{CristianoLaMagna2021, Poate1982}. It is nowadays exploited in several technologies, mostly due to its ultra-low thermal budget and its numerous control knobs (light wavelength and polarization, pulse duration, fluence, repetition rate, beam extension), which can be flexibly tuned to target specific functionalities, while handling the evergrowing complexity of nanosystems.

In the context of group IV elemental and compound semiconductor processing, pulsed-LA applications are ubiquitous \cite{CristianoLaMagna2021, Palneedi2018, Aktas2021review}. These include fabrication of poly-Si thin-film transistors \cite{Baeri1979, Privitera2007, Sameshima1986}, ultrashallow device junctions \cite{Privitera2007, Privitera2000, Vivona2022, AlvarezAlonso2022}, efficient contacts by silicidation \cite{Rascuna2019}, explosive crystallization \cite{Lombardo2018, AcostaAlba2021, Sciuto2020}, strain, defect \cite{Dagault2020, JohnsonII2022} and dopant engineering \cite{Mannino2005, Monflier2021, Daubriac2021, Huet2017, Palleschi2021}. Localized heating minimizes the risk of damaging sequentially integrated components of monolithic 3D devices \cite{Tabata2022, Salahuddin2018, Brunet2018, Fenouillet2021, Chery2022}. In optoelectronics pulsed-LA is a key-process for fabricating poly-Si displays \cite{Fortunato2012, Voutsas2003, Im2012, CoherentExcimer}, thin metal-oxides \cite{Palneedi2018}, pure-carbon electrodes for touch screens or solar cells \cite{Stock2020}, hyper-doped semiconductors for near-infrared photodetectors \cite{Mailoa2014}. It also allows strain, composition and morphology engineering of fiber-based photonic devices \cite{Aktas2021review} and fabrication of heavily-doped superconducting silicon for monolithic quantum device integration \cite{Bustarret2006, Chiodi2014, Daubriac2021, Hallais2023}.
Despite all these applications, understanding the ultrafast non-equilibrium kinetics of the liquid/solid interface in early stages of the process and correlating it to the post-irradiation morphology and properties is challenging. This is because any experimental characterization \cite{CristianoLaMagna2021}, no matter how accurate, can only access the final state of the system. Observations would indeed require in-situ, atomically-resolved and real-time capabilities well beyond those of modern electron-microscopy \cite{Alcorn2023} or atom-probe facilities \cite{Gault2021}. For this reason, computer simulations are nowadays indispensable for both fundamental studies and technological exploitation of LA.

LA simulations are usually deployed by self-consistently solving the electromagnetic interaction and heat diffusion problem in the irradiated system using a continuum description of its phase changes \cite{Huet2020, Wheeler1993, LaMagnaPhaseField2004, Aktas2021, Lin2014}. LA process parameters can be explored and fine-tuned in {\textmu}m-sized geometries, with the aid of computational libraries that use finite-element-methods (FEM) to solve the underlying coupled partial differential equations. However, continuum models cannot capture local nanoscale changes in the annealed materials with atomistic resolution. The latter may be a critical factor, especially for compound materials with complex 3D geometries or phase diagrams, crystal-orientation-dependent kinetics and defects, like stacking faults, which can affect the regrowth. Polymorphic solidification may also introduce structural disorder in the form of intermixed stacking motifs (e.g., cubic, hexagonal) \cite{Sadigh2021, Chen2020, Luo2022}. These phenomena can significantly alter the post-LA morphology and composition, with important consequences on device quality and performance. To ensure the appropriate process design and optimization, a simulation tool should be able to model the complex interplay between laser-matter interactions, the molten phase non-equilibrium kinetics and all possible atomic-scale structural transformations, while requiring the least amount of computational resources.

In this work we present a multiscale computational methodology enabling simulations of LA processes with atomic resolution. It is based on the local self-consistent coupling of a state-of-the-art {\textmu}m-scale FEM code with a Kinetic Monte Carlo on super-Lattice (KMCsL) code, able to simultaneously model atoms in the cubic and hexagonal crystal phases. Such multiscale approach enables atomistic modeling of extended defects, shape changes, composition and stack adjustments affecting the laser-annealed material up to hundreds of nm below the surface, while exchanging information between FEM and KMCsL at a ns pace. In this way, it not only overcomes the limits of purely continuum-based tools, but also those of other hybrid FEM-KMC approaches, which either lack self-consistent information exchange between the two frameworks \cite{Fisicaro2012} or are limited to defect-free LA simulations of silicon without any super-lattice formulation \cite{Calogero2022}.
In particular, we demonstrate the method by focusing on ultraviolet ns-pulsed LA processes of SiGe, an alloy with composition-dependent electronic and optical properties \cite{Olesinski1984, Wang2021, Li2022} increasingly relevant to future nanoelectronic \cite{IRDS2022, Xiang2006, Muller2022, Wang2019, Brunet2018, Koo2016, Wind2022, Li2021}, thermoelectronic \cite{Basu2021}, optoelectronic \cite{Aktas2021review, Fadaly2020, Coucheron2016} and quantum technologies \cite{McJunkin2022, Scappucci2020, Giustino2020, Losert2023}. The multiscale methodology provides unique atomistic insights on the complex and ultrafast morphological, compositional and structural transformations of SiGe during laser irradiation \cite{Dagault2019, Dagault2020, JohnsonII2022, Gluschenkov2018}, giving invaluable support to process engineers aiming at the exploitation of this material's full potential.

\section{Results}
\begin{figure*}[t]
  \centering
  \includegraphics[width=0.98\linewidth]{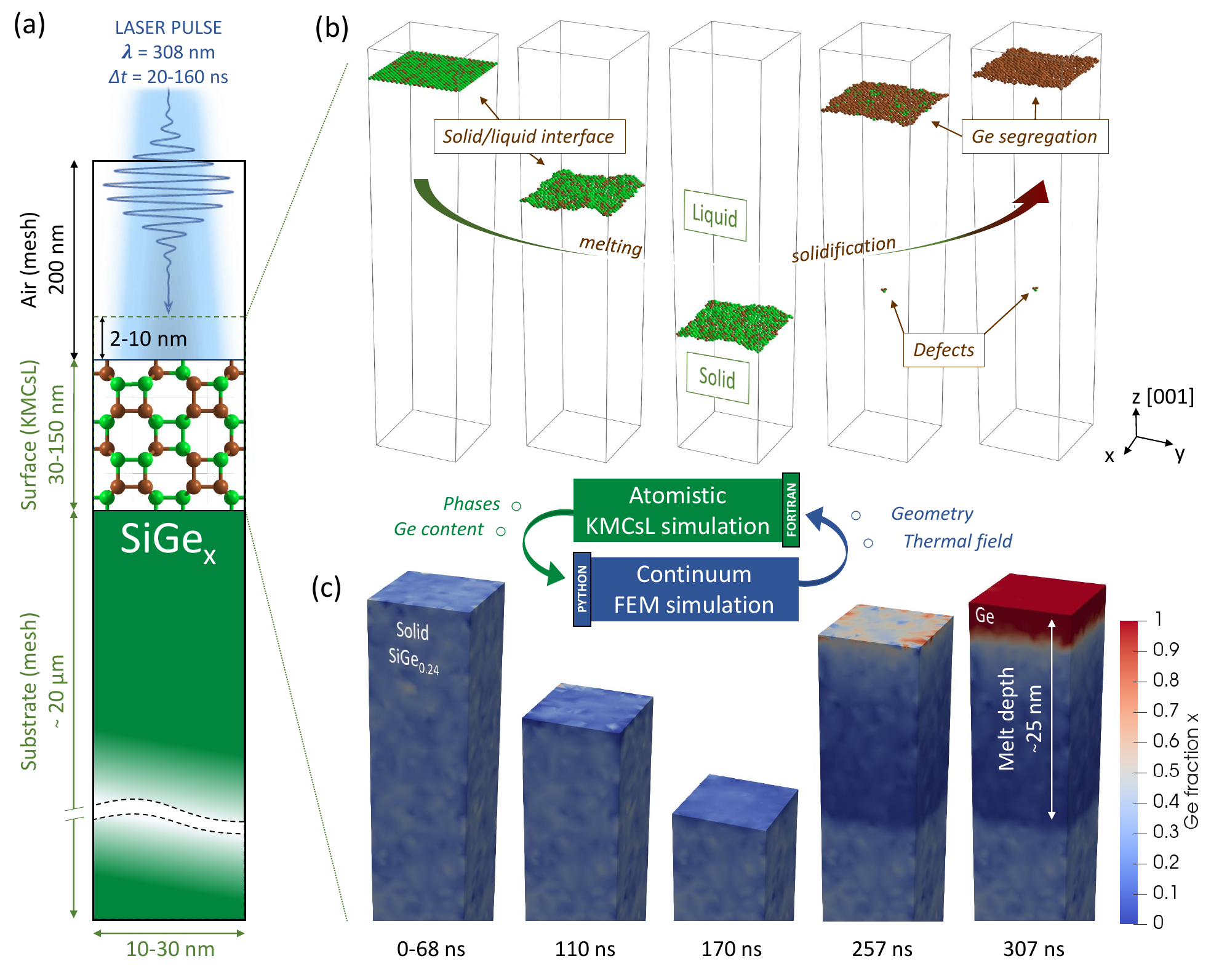}
  \caption[Schematics of the FEM-KMCsL approach.]{Schematics of the FEM-KMCsL multiscale approach applied to a SiGe \hkl(001) surface. (a) Sketch of the simulation framework. Typical simulation box dimensions are also indicated. (b) Liquid-solid interface in the KMCsL box at various instants during melting and solidification. Solid under-coordinated Si and Ge atoms (green and brown, respectively) identify the interface. (c) Ge content in solid SiGe in the FEM model, visualized at the same instants. Pure Si (x=0) regions in blue and pure Ge (x=1) in red.}
  \label{fig:fig2}
\end{figure*}
\begin{figure*}[t]
  \centering
  \includegraphics[width=0.98\linewidth]{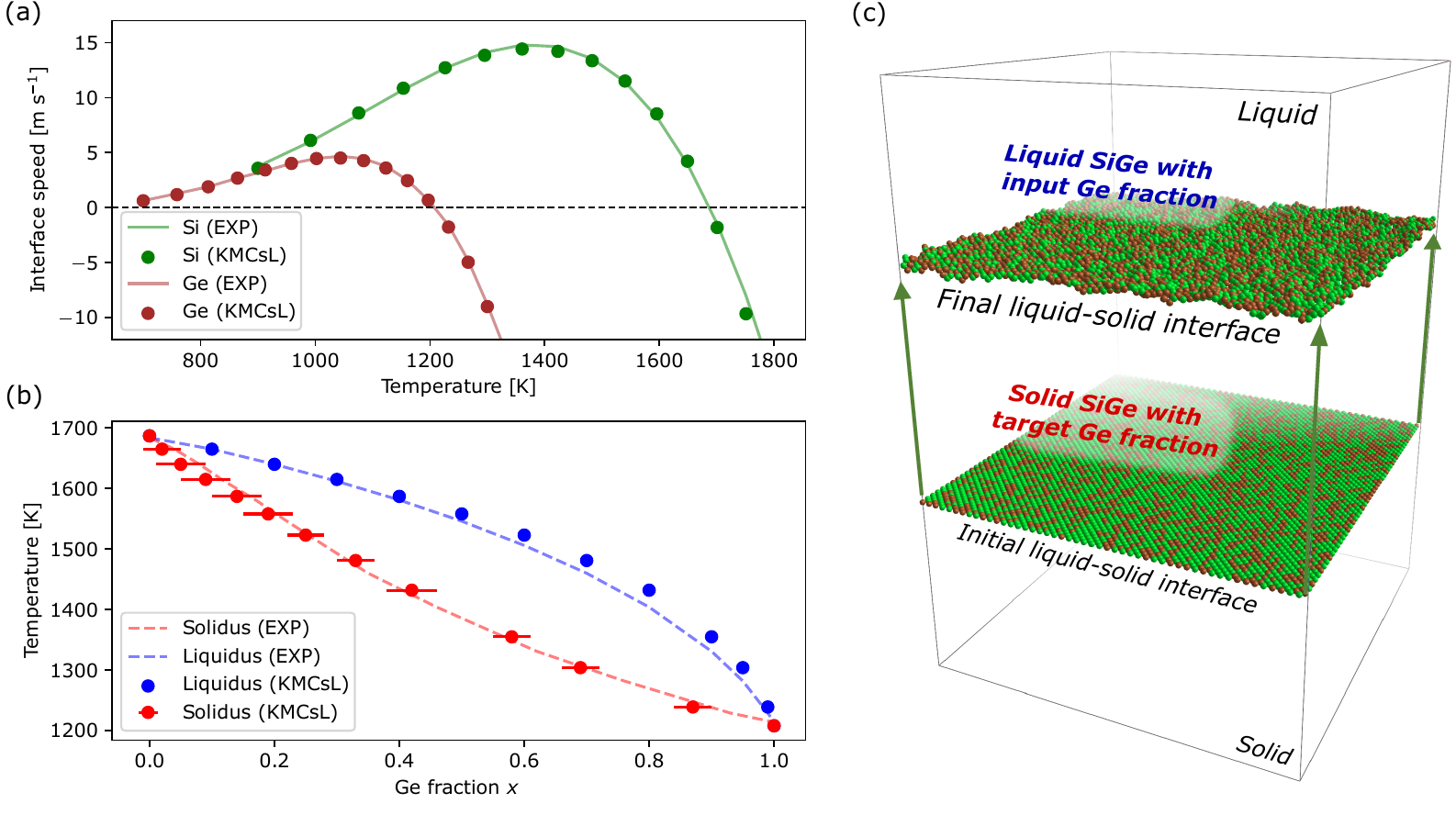}
  \caption[KMCsL calibration.]{KMCsL calibration for SiGe solid-liquid phase transitions. (a) Calibrated results for the solid-liquid interface velocity in pure Si and pure Ge as a function of temperature (markers), in comparison with the respective Fulcher-Vogel profile (lines). (b) Calibrated results for the SiGe phase diagram (markers), along with the expected phase diagram (dashed lines). The horizontal bars reflect spatial compositional variations in the solidified layer. (c) Schematics of the strategy followed for phase diagram calibration.  The superimposed snapshots show initial and final states of a calibrated simulation for $T\approx1410$ and $x_L\approx0.8$, resulting in $x_S\approx0.43$. Solid under-coordinated Si and Ge atoms (green and brown, respectively) identify the interface.}
  \label{fig:fig1}
\end{figure*}

\subsection{Multiscale FEM-KMCsL coupling}
\label{sec:A}

Our LA simulations are based on a multiscale coupling between FEM and KMCsL solvers, which exchange information in a self-consistent loop at time steps $\Delta t<1$ ns throughout the simulation. Besides providing atomistic insights, this approach ensures higher accuracy compared to pure phase-field or enthalpy formalisms \cite{CristianoLaMagna2021}, as the latent heat exchanged at every $\Delta t$ is computed by direct integration of the volume subjected to a phase transition during each KMCsL step. The multiscale procedure is hereby described. After setting up the appropriate 3D mesh for a system with desired size and composition, the FEM calculation begins. The laser-induced heat source and temperature field $T(t,\textbf{r})$ within the irradiated material are self-consistently calculated, by solving the Maxwell's and Fourier's partial differential equations. As the system absorbs energy from the laser pulse, following its power density modulation, the surface temperature rises until local melting occurs. This initiates the feedback coupling with KMCsL, which models atom-by-atom the concerned system subregion. The following steps are then iterated every $\Delta t$ over the whole pulse duration: \begin{itemize}\item $T(t,\textbf{r})$ is interpolated into the dense KMCsL super-lattice and defines melting/solidification events probabilities; \item KMCsL simulates the evolution of the solid/liquid (S/L) interface for a time $\Delta t$ with the established probability table, capturing atomic-scale structural adjustments, lattice faceting, vacancies, extended defects, polymorphic solidification and species redistribution; \item the S/L volumes and the local species concentrations are updated in the mesh based on the KMCsL results and affect the $T(t,\textbf{r})$ calculation in the subsequent FEM cycle. \end{itemize} The above three steps are iterated until all previously-melted atoms resolidify. Thereafter the FEM-KMCsL communications stop and the FEM model is left to cool. 
\Fref{fig:fig2} schematically illustrates a typical FEM-KMCsL simulation box (characteristic sizes used in this work are also indicated) for modelling pulsed-LA of a flat \sige{0.76}{0.24} \hkl(001) surface. \Fref{fig:fig2}b shows the solid atoms at the S/L interface in the KMCsL-modelled subregion, at various instants of a simulation assuming a XeCl excimer ($\lambda=308$ nm) 160 ns laser pulse, with 0.75 J cm$^{-2}$ energy density and a $\Delta t=0.25$ ns. After the initial heating stage up to $T_M(x=0.24)\approx1573$ K, the interface goes deep into the material (roughly 25 nm), keeping a roughness of a few nm. Then it rapidly ascends as $T$ decreases, solidifying a SiGe layer with graded Ge content and a Ge-rich surface, due to non-equilibrium species partitioning in the alloy \cite{Brunco1995}. The corresponding solid-phase regions in the 3D FEM mesh at the same instants are reported in \Fref{fig:fig2}c, with colours highlighting Ge segregation.

More details on the multiscale implementation are reported in Methods (see also Supplementary Note S1, Figure S1 and Movie S1).

\subsection{Calibration of KMCsL for SiGe S/L interface}
\label{sec:B}

In the KMCsL model of a partially-melted system, the evolution of the S/L interface is governed by the balance between solidification and melting events with $T$-dependent Arrhenius-like probabilities (see Methods). To ensure reliable LA simulations, it is crucial to calibrate the KMCsL event probabilities so that they reproduce the correct S/L interface kinetics over a wide range of temperatures. In case of SiGe alloys, this calibration is carried out in two steps. The first, following the strategy of Ref. \cite{Calogero2022}, consists in reproducing the Fulcher-Vogel curves of pure Si and Ge systems \cite{Stiffler1992, Lombardo2018}, i.e., the S/L interface velocity as a function of $T$. We do this by initializing solid Si and Ge surfaces surmounted by an infinite liquid reservoir and performing a sequence of KMCsL simulations for a wide range of $T$ around $T_M$, always assuming $T$ uniformity in the simulated box. The KMCsL parameters are then fine-tuned to yield the expected interface velocities, as shown in \Fref{fig:fig1}a (calibrated values in Supplementary Table S1-3).
The second step, since no Fulcher-Vogel relation holds for SiGe alloys, consists in calibrating the KMCsL event rates involving mixed Si-Ge bonds on the SiGe phase diagram \cite{Olesinski1984} (dashed lines in \Fref{fig:fig1}b), which describes the $T$-dependent composition $(x_S, x_L)$ of solid and liquid phases at equilibrium, when the melting/solidification process occurs very slowly. This is achieved by setting up slightly under-cooled KMCsL simulations around a given $(T, x_L)$ point in the phase diagram and tuning the parameters until the solidified SiGe layer roughly matches the expected $x_S$ (more details in Supplementary Note S2). For example, \Fref{fig:fig1}c shows two snapshots of the S/L interface at the beginning and at the end of a well calibrated simulation assuming $(T\approx1410, x_L\approx0.8)$. This predicts a solid layer with average $x_S\approx0.43$, which is in line with the phase diagram and hence confirms the reliability of the calibration at the considered $T$. The simulated results spanning the entire phase diagram from Si-rich to Ge-rich situations are reported in \Fref{fig:fig1}b.

\subsection{Validation of multiscale LA simulations for relaxed/strained SiGe}
\label{sec:D}
\begin{figure*}[t]
  \centering
  \includegraphics[width=0.95\linewidth]{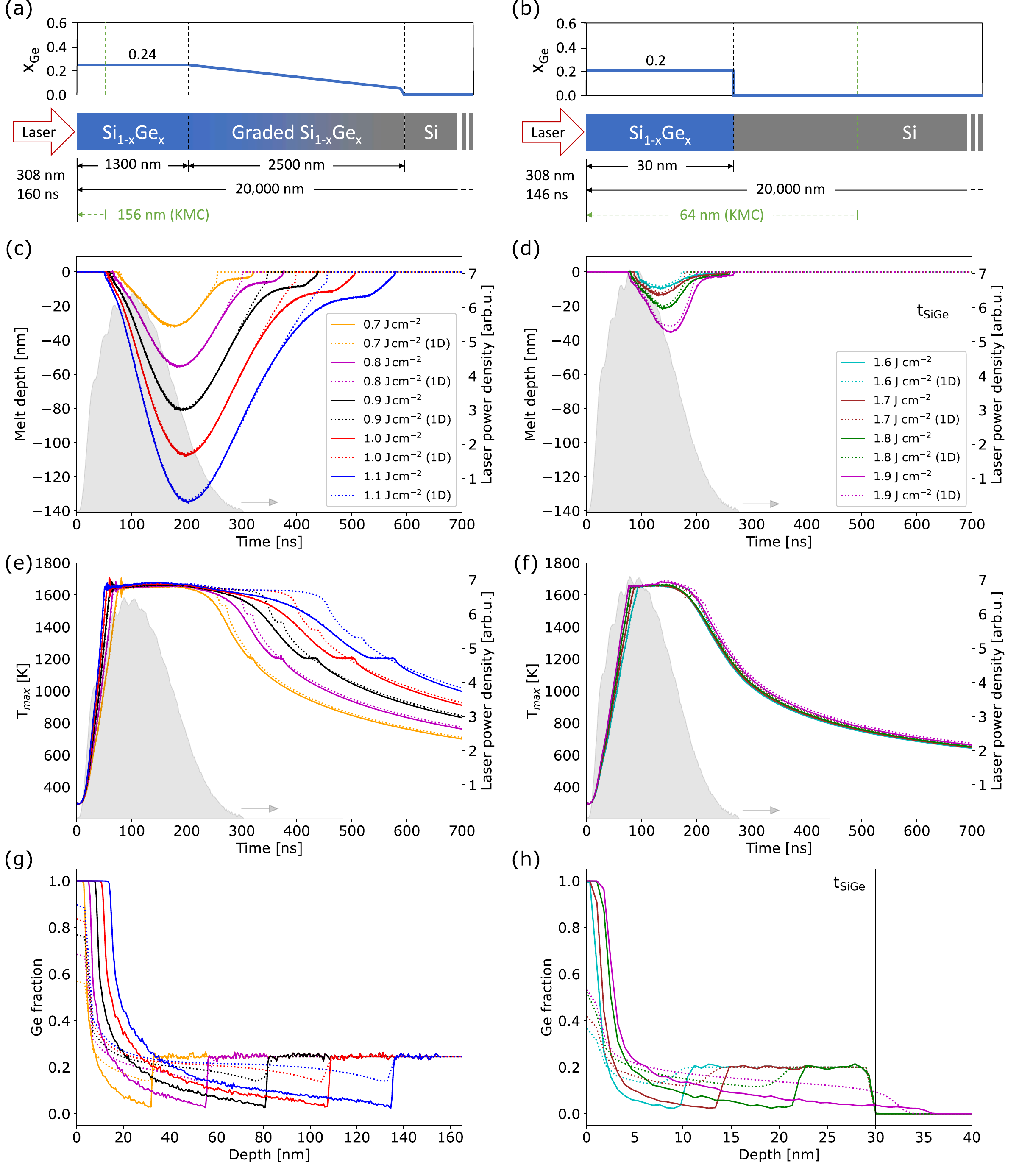}
  \caption[Validation against 1D phase field.]{Comparison between atomistic FEM-KMCsL (solid lines) and non-atomistic phase-field simulations (dotted lines). (a,b) Pulsed-LA simulation setup for relaxed (left) and strained SiGe (right) with thickness $t_{\rm SiGe}=30$ nm, including system sizes and initial composition. (c-d) Time variation of melt depth, (e-f) maximum temperature in mesh and (g-h) post-anneal Ge profiles obtained for various laser energy densities. The laser pulse shape (filled grey areas) and power density (right axes) are also reported.}
  \label{fig:fig3}
\end{figure*}
Here we show the results of FEM-KMCsL simulations for relaxed and 30 nm-thick strained SiGe\hkl(001) layers epitaxially grown on Si. To check the consistency of the method and validate it, we compare these results with those of 1D non-atomistic simulations based on a state-of-the-art FEM-phase-field formulation (see Methods), considering pulsed-LA processes with various laser energy densities and pulse durations (160 ns for relaxed, 146 ns for strained). The simulation settings (optical/thermal parameters, initial Ge profile, laser properties) in the 1D purely continuum and the 3D FEM-KMCsL multiscale frameworks are identical, except for the mesh dimensionality and the formalism describing phase transitions (phase-field with smooth S/L interface in one case, KMCsL with atomically sharp S/L interface in the other). 

The initial Ge profiles and process conditions are reported in \Fref{fig:fig3}a-b. \Fref{fig:fig3}c-d shows that the two methodologies yield almost identical results concerning the general melt depth profile over time, with KMCsL yielding melting/solidification velocities and maximum melt depths in remarkably good agreement with phase-field simulations. Contrary to the latter, the FEM-KMCsL approach can track the interface evolution up until the last solidification event and, as a result, it can reproduce the expected slowdown of the solidification front due to the gradual Ge incorporation. \Fref{fig:fig3}e-f shows the variation of maximum temperature $T_{\rm max}$ in the mesh over the same time interval. Both models predict an overall similar trend, with the expected change in slope at the onset of melting and almost overlapped cooling tails after complete solidification. The observed $T_{\rm max}$ plateaus at the end of solidification are also related to Ge segregation, as those observed in \Fref{fig:fig3}c-d. Their position and dependence on energy density differ between the two models, in accordance with the final profiles of Ge concentration over depth reported in \Fref{fig:fig3}g-h. Like all other noticeable deviations in \Fref{fig:fig3}, this can be attributed to intrinsic differences between the two models (more details in Supplementary Note S3).
Overall, these results confirm the internal consistency of the multiscale methodology for a wide range of process conditions and demonstrate that the phenomenon of Ge segregation is qualitatively captured in both relaxed and strained SiGe.

\begin{figure}[t]
  \centering
  \includegraphics[width=0.999\linewidth]{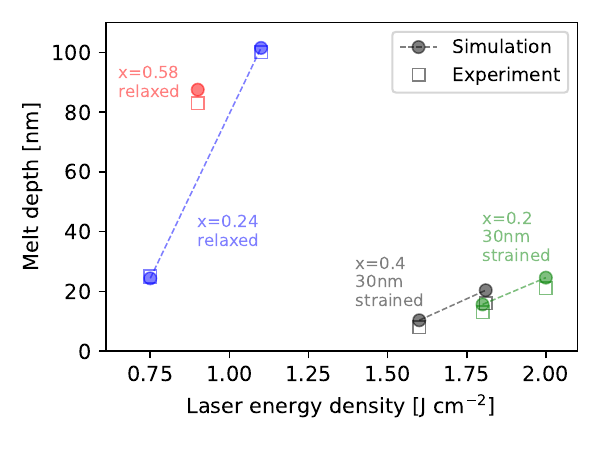}
  \caption[Comparison with experiments.]{Comparison between maximum melt depths simulated with FEM-KMCsL (circles) and experimental measurements (squares), for various energy densities and Ge fractions. Relaxed (strained) samples are irradiated with a 308 nm, 160 ns (146 ns) pulse. Dashed lines guide the eye through processes on the same sample.}
  \label{fig:fig4}
\end{figure}
Further validation is provided by the results in \Fref{fig:fig4}. These demonstrate that the maximum melt depths obtained with FEM-KMCsL simulations are also in good agreement with experimental measurements (see Methods). In particular, here we consider processes with 160 ns pulses for relaxed \sige{0.76}{0.24} and \sige{0.42}{0.58}, and with 146 ns pulses for strained 30nm-thick \sige{0.8}{0.2} and \sige{0.6}{0.4}. The maximum melt depths and Ge fraction for the relaxed samples were extracted from Energy Dispersive X-ray spectroscopy measurements performed after the irradiation. Data for strained samples is taken from Refs. \cite{Dagault2019, Dagault2020}.

\subsection{LA simulations with stacking faults}
\label{sec:E}
\begin{figure*}[t]
  \centering
  \includegraphics[width=0.98\linewidth]{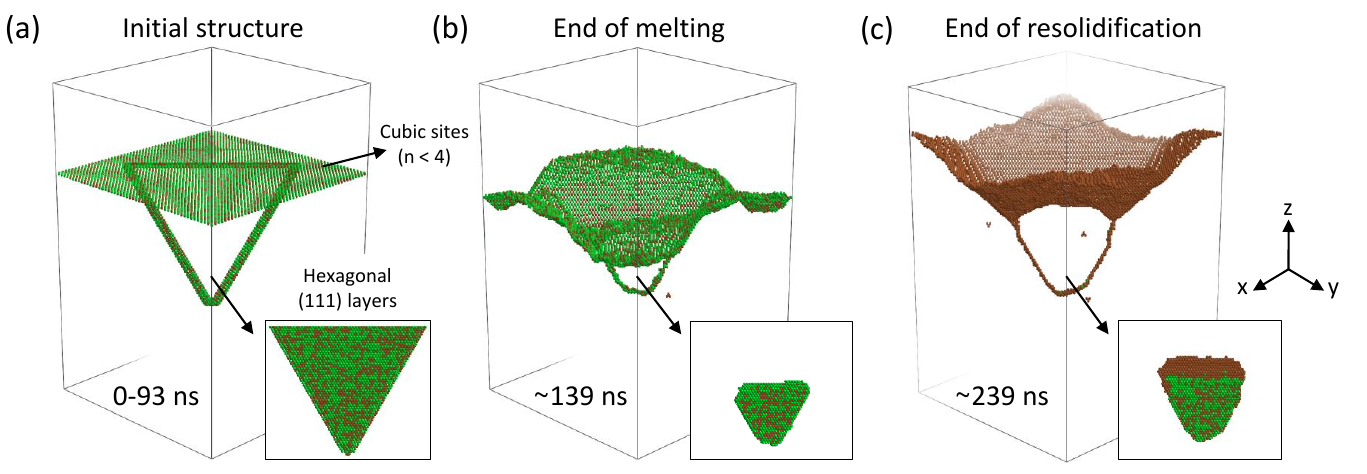}
  \caption[LA process with stacking fault.]{Multiscale LA simulations (308 nm, 146 ns, 1.3 J cm$^{-2}$) for strained \sige{0.6}{0.4} on Si with pre-existing triple stacking fault. Undercoordinated cubic solid atoms in the KMCsL box are shown at three instants during the process: (a) before melting, (b) at the maximum melt depth and (c) at the end of solidification. Hexagonally stacked atoms (regardless of coordination) at all instants are shown in the insets.}
  \label{fig:fig5}
\end{figure*}

The KMCsL formulation allows to study the impact of extended stacking defects on the final morphology of laser-annealed materials (see Methods). As an example, in \Fref{fig:fig5} an LA process for strained 30nm \sige{0.6}{0.4} on Si is considered (146 ns, 1.3 J cm$^{-2}$), where a $\sim$10 nm-deep triple stacking fault exists in the sample prior to laser irradiation. This introduces three hexagonally stacked \hkl(111) atomic layers into the cubic SiGe structure. All cubic undercoordinated surface atoms in the KMCsL box before melting are shown in \Fref{fig:fig5}a, along with the bulk ones enclosing the \hkl(111) layers. The S/L interface at the maximum melt depth and after full solidification are shown in \Fref{fig:fig5}b-c (see also Supplementary Movie S2). We find that part of the defect is melted along with 7-8 nm of SiGe, without significant impact on the melting kinetics, and that a strongly inhomogeneous solidification is triggered by the unmelted hexagonal sites. Liquid atoms in direct contact with them indeed solidify much slower than the others, favouring \hkl{111} faceting of the S/L interface (see \Fref{fig:fig5}a-c). Segregation is observed in both cubic and hexagonal crystal phases (see insets in \Fref{fig:fig5}). For higher energy densities, the defect is fully melted and a purely cubic phase planar solidification occurs as usual. 

\subsection{LA simulations with nanostructured and constrained geometries}
\label{sec:F}

\begin{figure*}[t]
  \centering
  \includegraphics[width=0.95\linewidth]{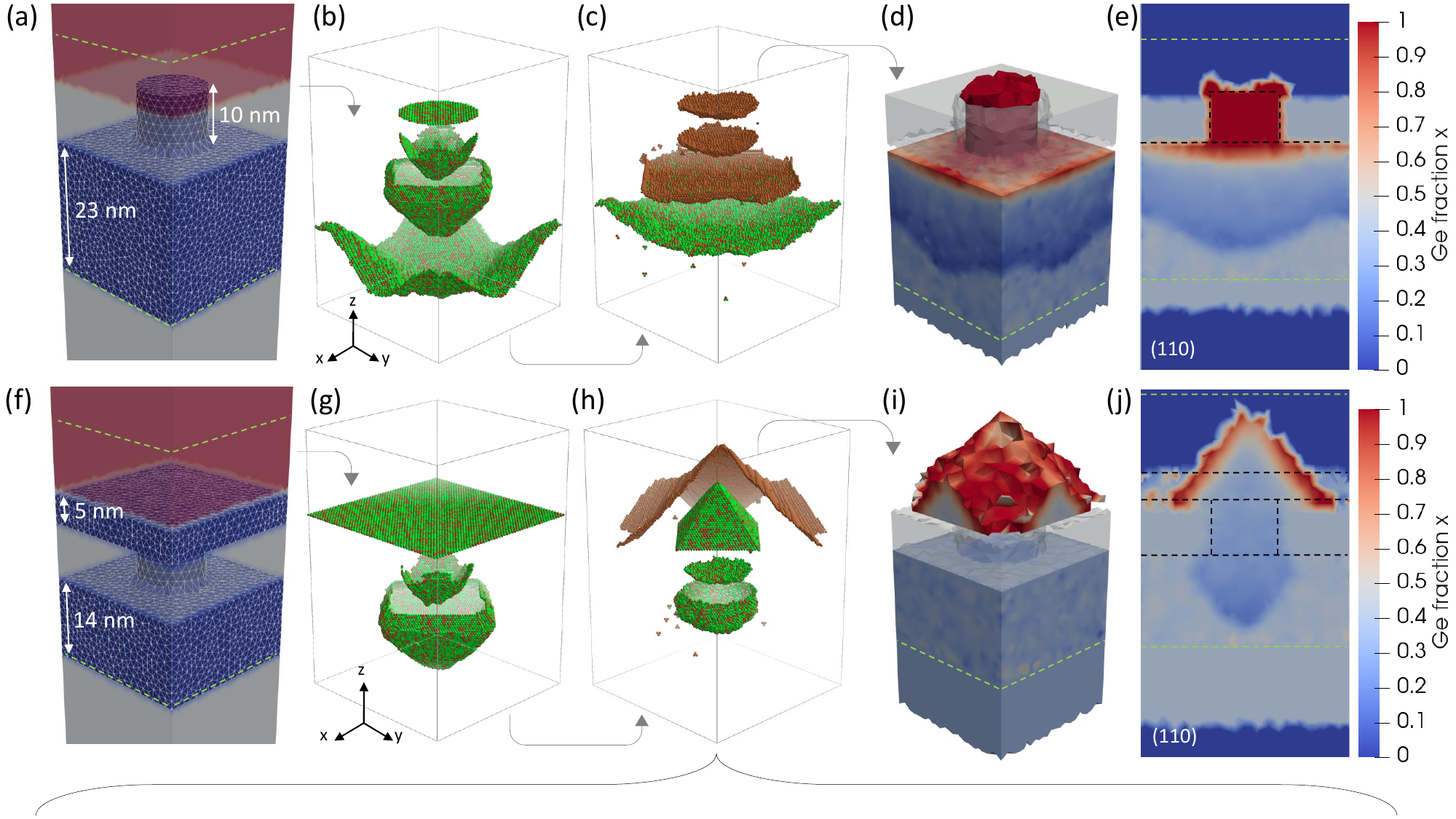}
  \includegraphics[trim=0 100 0 0, clip, width=0.95\linewidth]{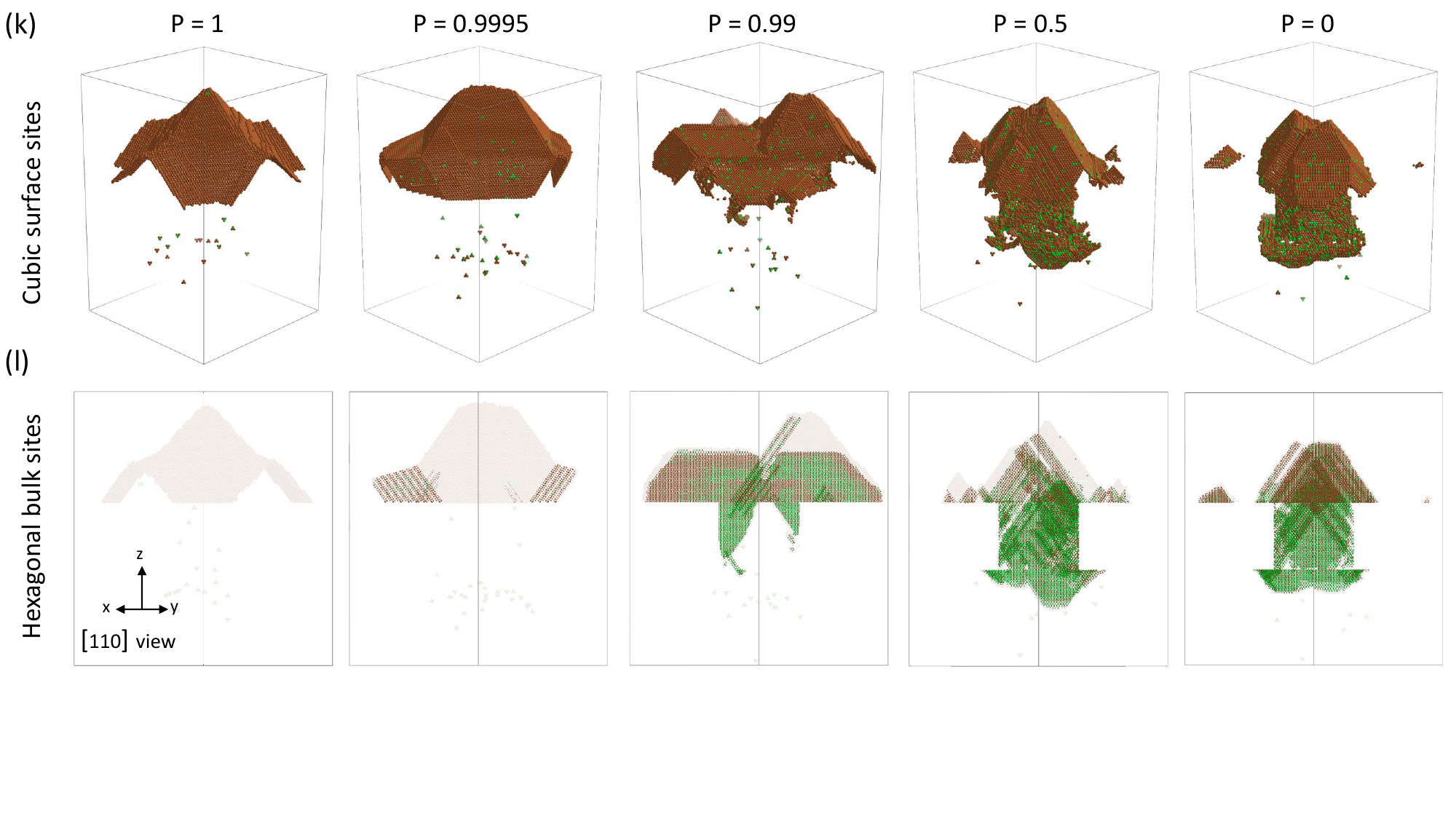}
  \caption[Hybrid simulation of SiGe oxide-embedded NWs.]{
  (a-e) Pulsed-LA simulation (308 nm, 22 ns, 0.95 J cm$^{-2}$) of 30nm strained \sige{0.6}{0.4} with 9nm-large and 10nm-high \sige{0.6}{0.4} NWs on top, embedded in non-melting SiO$_2$. (a) Input FEM periodic mesh. KMCsL-coupled SiGe regions are shown in blue, air in red, SiO$_2$ and non-KMCsL-coupled regions in grey. The KMCsL cell estension along $z$ is indicated with dashed green lines. (b) Overlapped selected snapshots of the liquid-solid interface in the KMCsL box at various instants during melting and (c) solidification. Green (brown) spheres indicate Si (Ge) atoms. (d) 3D view and (e) \hkl(110) cut-plane of the final Ge distribution in the FEM mesh. Regions outside the KMCsL cell (below the green dashed line) appear uniformly coloured because no KMCsL mapping occurs therein. The initial surface morphology is indicated by dashed black lines. (f-j) Simulation at 1.2 J cm$^{-2}$ energy density for the same system as above, including a 5nm-thick \sige{0.6}{0.4} capping layer. (k) Cubic undercoordinated sites in the KMCsL box at the end of the simulations performed at the same conditions as (f-j), but with different polymorphic solidification probabilities $P$. (l) Hexagonally stacked sites (regardless of coordination) from the simulations in (k), viewed along the \hkl[110] direction. Cubic sites from (k) are redrawn in semi-transparency.}
  \label{fig:fig6}
\end{figure*}

The previous example reveals another important feature of KMCsL, i.e., the crystal-orientation-dependent kinetic evolution of the S/L interface. Such an atomistic feature is essential to model LA of SiGe systems with nanostructured and/or constrained geometries, which often involve reshaping and faceting of both liquid and solid volumes throughout the process.
An example is illustrated in \Fref{fig:fig6}a-e. It considers an LA process (22 ns, 0.95 J cm$^{-2}$) of a SiGe system similar to those used in vertical nanostructured channel arrays \cite{Muller2022}, namely a 30nm-thick strained \sige{0.6}{0.4} on Si with a 9nm-large and 10nm-high \sige{0.6}{0.4} nanowire on top. The latter is embedded in SiO$_2$, which does not melt during the irradiation and therefore represents a geometrical constraint for the evolving S/L interface. An energy density of 0.95 J cm$^{-2}$ is chosen to keep melting within the KMCsL box, which is $27\times 27\times 41$ nm$^3$ and includes $\sim$23 nm of the SiGe layer, the nanowire, the oxide and $\sim$8 nm of air (see green dashed lines in \Fref{fig:fig6}a). \Fref{fig:fig6}b-c shows snapshots of the S/L interface at various instants during melting and solidification (see also Supplementary Figure S3 and Movie S3). The circular nanowire tip exposed to the laser rapidly absorbs heat and melts all the way down to the bottom of the oxide. In this process, the shape of the S/L interface is already \hkl{111}-faceted. After complete melting of the nanowire, the nanodroplet reshapes into a half-octahedron below the oxide, which then coalesces with its periodic images, giving rise to a rough liquid layer, similar to what occurs in simulations of Si LA processes assuming inhomogeneous nucleation \cite{Calogero2022}. Thereafter, the S/L interface flattens and moves towards the initial surface level. While constrained by the oxide, Ge segregation occurs, causing a total transformation of the initial SiGe nanowire into a pure Ge nanowire. The final Ge distribution in the FEM mesh is illustrated in \Fref{fig:fig6}d and in \Fref{fig:fig6}e as a \hkl(110) cut-plane. These figures show the rough shape of the interface at the maximum melt depth and highlight a slight tendency towards solidification of the nanowire shell, before its core (also noticeable in the KMCsL snapshots). 

By initializing the above simulation with an additional 5 nm-thick \sige{0.6}{0.4} capping layer (see \Fref{fig:fig6}f), we trigger solidification on top of the nanowire/oxide region and give rise to more pronounced solid-phase reshaping effects. This time we included $\sim$14 nm of SiGe, the nanowire, the oxide, the capping layer and $\sim$11 nm of air within the KMCsL box, and used 1.2 J cm$^{-2}$ as energy density (enough to avoid coalescence of molten nuclei). The kinetic evolution in \Fref{fig:fig6}g-h (see also Supplementary Movie S4) and the final Ge distributions in \Fref{fig:fig6}i-j reveal that the S/L interface initiates solidification with a non-planar shape and assumes a highly symmetrical \hkl{111}-faceted pyramidal shape as it emerges above the oxide. This solid seed gradually expands and partially coalesces with its periodic replicas, while concurrently segregating Ge.

\subsection{LA simulations with polymorphic solidification}
\label{sec:G}

As a further demonstration of the potential of KMCsL for LA simulations, in \Fref{fig:fig6}k-l we report on the results of the previous simulation obtained while allowing for polymorphic cubic-hexagonal stacking transitions during solidification. \Fref{fig:fig6}k depicts the cubic undercoordinated KMCsL sites at the end of LA simulations performed by varying the probability $P$ of switching the stacking order (see Methods). \Fref{fig:fig6}l shows all hexagonally stacked atoms superimposed to the cubic ones (semi-transparent), viewed along the \hkl[110] direction to highlight the presence of \hkl{111} atomic layers. We find significantly intermixed stacking motifs, even for very small probabilities. A high concentration of stacking faults (both single and triple) characterizes the oxide-embedded region, suggesting a clear correlation between confinement and stacking disorder. Noteworthy, the pyramidal shape of the final surface is quite robust against polymorphic disorder (see also Supplementary Figure S3 and Movie S5-S6).

\section{Conclusions}
We have presented a new multiscale approach to model LA processes of group IV materials and alloys, including complex 3D shape modifications, liquid and solid-phase faceting, species redistribution, stacking disorder and extended defects. It is based on the self-consistent combination of a continuum FEM-based solver for light-matter interaction and thermal diffusion with a KMCsL code. The latter simulates the kinetic evolution of liquid-solid interface and lattice defects in a local region of the material with atomic resolution, enabling studies so far inaccessible to purely continuous simulation approaches. 
In particular, we have described the theoretical background and computational implementation of the methodology in light of its application to the \sige{1-x}{x} alloy, which represents one of the most promising candidates for 3D sequentially integrated devices \cite{Wang2019, Brunet2018, Koo2016, Wind2022}, spin-qubits \cite{Losert2023}, Gate-All-Around transistors \cite{Muller2022, Xiang2006, Li2021} or even direct-bandgap light emitters \cite{Fadaly2020}. The method was validated by comparing simulations for both relaxed and strained SiGe with 1D phase-field results and experiments. It quantitatively reproduces the same melt depth profiles and qualitatively captures laser-induced Ge redistribution. KMCsL has the advantage of avoiding the typical numerical instabilities of approaches based on phase-field, especially at the onset and the end of melting. 
The code was applied to simulate pulsed-LA processes of blanket and nanostructured SiGe systems, including effects of extended defects and geometrical constraints. The possibility of studying the impact of extended defects and polymorphic solidification was demonstrated, and a clear correlation between bulk structural disorder and post-irradiation surface morphology was observed.

Importantly, the methodology is implemented into an open-source versatile tool which offers several opportunities in terms of potential generalizations. The unique KMCsL super-lattice framework, enabling the coexistence of multiple crystal arrangements in the same simulation box, is readily applicable to other elemental or compound group IV semiconductor with sp$^3$ bond symmetry, e.g., Si, Ge or SiC \cite{Calogero2022, Fisicaro2020}. By tailoring the crystal symmetries, it could be generalized to other binary alloys (e.g., GeSn) \cite{Wirths2015}, compound semiconductors (e.g., GaAs, AlGaAs) or polymorphic metal/semiconductor systems (e.g., NiSi, PtSi). Future KMCsL developments may broaden the kinetic landscape by including liquid-phase diffusion events and strain relaxation events. 
With properly calibrated FEM optical and thermal parameters, lasers with different wavelengths could be studied. Continuous-wave and scanning LA processes could also be investigated by adjusting the input profile of laser power density. 

Framing the FEM-KMCsL strategy into a broader multiscale perspective, one may envision advanced coupling with other \textit{ab initio}, molecular dynamics or transport simulation tools, e.g., to account for strain relaxation and interactions between extended defects \cite{Barbisan2022} or investigate the impact of ultrafast processing on device components \cite{Adetunji2021}. A similar multiscale approach could be used to study processes where other physical variables govern the atomic kinetics (e.g. strain, charge, polarization, magnetization), or where phase transitions are triggered by different ultrafast external stimuli (e.g. electric, magnetic or strain perturbations) \cite{Yang2022}. This could provide interesting insights in various research areas, from silicidation \cite{Rascuna2019} to multiferroics \cite{Liou2020} or phase-change resistive-switching materials for neuromorphic computing and high-speed photonic-based devices \cite{Shen2021, Wang2020}.

\section*{Methods}
\subsubsection*{Multiscale implementation}
The multiscale simulation tool developed in this work is distributed as part of the open-source \mulskips simulation package \cite{mulskips, LaMagna2019}. This includes a core KMCsL code built on a peculiar super-lattice framework which enables simultaneous modelling of cubic and hexagonal crystal phases in the same simulation cell. Such a functionality is critical for LA simulations of multi-element systems including non-ideal stacking and polymorphism. With the appropriate particle/event definition and calibration, it can also simulate epitaxial growth by physical or chemical vapour deposition \cite{Fisicaro2020}. The KMCsL model, coded in Fortran, is internally coupled to a FEM-based solver coded in Python with the \dolfin interface of the \fenics computing platform \cite{Alnaes2015} (the same solver is used for the  benchmark 1D phase-field simulations). The \pymulskips Python library, distributed with \mulskips, manages all simulation workflows and includes an I/O interface coupling the KMCsL simulator to the FEM solver and external Technology Computer-Aided Design (TCAD) tools. In particular, \emph{ad-hoc} Application Programming Interfaces are implemented to manage the multi-process shared-memory execution of simulations via F2Py sockets \cite{f2py1, f2py2}. This ensures a real-time communication of all relevant geometrical and physical information between the different simulation frameworks. By allowing a single KMCsL process to run over the entire simulation, it also enables a continuous tracking of species position and bonding configurations, which is crucial for simulating the evolution of extended defects or polymorphic domains across consecutive FEM-KMCsL cycles (more details in Supplementary Note S1).

\subsubsection*{KMCsL model}
The KMCsL model is defined on a dense cubic super-lattice able to accommodate both cubic and hexagonal diamond lattices as sub-lattices. The super-lattice constant is $a_{\rm KMCsL}\equiv a/12\equiv l/\sqrt{27}$, with $a$ the diamond lattice constant and $l$ its nearest neighbour distance (0.543 nm and 0.235 nm for Si, respectively). This definition makes it readily applicable to any elemental, compound and alloy material with sp$^3$ (tetrahedral) bond symmetry, such as Si, Ge or SiGe, including non-ideal stacking configurations. 
Each super-lattice site is marked as either solid or liquid site and can have coordination $n\leq4$ (for $n=4$ they are marked as bulk). In case of \sige{1-x}{x}, the two atomic species are randomly allocated in the lattice of the input structure, reflecting the user-defined Ge fraction $x$.  
To reduce calibration parameters and memory consumption, thus improving scalability, a real atomic occupancy is strictly considered only in the solid phase (the accuracy of this approach was already demonstrated for Si LA simulations \cite{Calogero2022}). In case of SiGe alloys, while the solid-phase Ge fraction can be described as a local time-dependent variable, $x_S\equiv x_S({\bf r}, t)$, the liquid-phase Ge fraction is averaged over the liquid volume at each ns-long KMCsL cycle $x_L\equiv x_L(t)$ (more details in Supplementary Note S3).  
In a partially-melted system, the kinetic evolution of the liquid-solid interface is governed by the balance between solidification and melting events. These are stochastically selected by a continuous time algorithm \cite{Calogero2022} and only involve under-coordinated ($n<4$) super-lattice sites. 
We note that no kinetics occurs in the bulk, no matter if solid or liquid. Diffusion events are not explicitly defined in the KMCsL framework, but they can be effectively reproduced by close melting/solidification events nearby the interface. 
The solidification and melting event rates are thermally activated and therefore obey Arrhenius-like expressions, with prefactors and exponents depending on temperature $T$ and bond coordination $n$ \cite{Calogero2022}. In the case of SiGe alloys, they also explicitly depend on the fraction $X^i$ of individual species in the liquid phase, with $X^i=x_L$ for Ge and $X^i=1-x_L$ for Si. In particular, the solidification (melting) event rate $\nu_{\rm LS}^i$ ($\nu_{\rm SL}^i$) for species $i=\textrm{Si,Ge}$ on a site with $n=n_{\rm Si}+n_{\rm Ge}$ solid neighbours is defined as:
\begin{align}
    \nu_{\rm LS}^i &= f^i(T) \cdot X^i \cdot \nu_0^i  \cdot {\rm exp}\left(-\frac{2 E_{\rm LS}^i (n)}{k_B T_M^i}\right) \label{eq:1}\\
    \nu_{\rm SL}^i &= \nu_0^i  \cdot {\rm exp}\left(-\frac{n E_{\rm SL}^i (n_{\rm Si}, n_{\rm Ge})}{k_B T}\right) \label{eq:2}
\end{align}
where $\nu_0^i$ is a species-dependent constant prefactor, $T_M^i$ is the melting temperature of species $i$ and $k_B$ is the Boltzmann constant. The solidification rate increases with $X^i$ and a damping term $f^i(T) = 1/2 [1+{\rm erf}((T-T_0^i)/A^i )]$, with $T_0^i$ and $A^i$ adjustable parameters, effectively models the reduction in solidification velocity under strong undercooling, as predicted by the Fulcher-Vogel expression \cite{LaMagnaPhaseField2004, Mittiga2000}. $E_{\rm LS}^i (n)$ is the $n$-dependent solidification energy barrier for species $i$, whereas $E_{\rm SL}^i (n_{\rm Si}, n_{\rm Ge})$ is the melting energy barrier, which is equivalent to a binding energy and hence depends on the number and identity of nearest neighbours. We note that the event rates $\nu_{\rm LS}^i\equiv\nu_{\rm LS}^i(t, {\bf r})$ and $\nu_{\rm SL}^i\equiv\nu_{\rm SL}^i(t, {\bf r})$ vary with time $t$ and lattice position ${\bf r}$ during the LA simulations. This stems from the fact that $T\equiv T(t, {\bf r})$ as well as $X^i\equiv X^i(t)$ are evaluated at every FEM-KMCsL cycle. All parameters in \Eref{eq:1} are determined by calibrating pure Si and Ge systems against their Fulcher-Vogel curves (\Fref{fig:fig1}a). The melting energy barriers, $E_{\rm SL}^i (n_{\rm Si}, n_{\rm Ge})$, are determined by calibrating SiGe against its experimental lens-shaped phase diagram (\Fref{fig:fig1}b). The precise expressions for the energy barriers and the detailed procedure for calibration are reported in Supplementary Note S2. 

Solidification at a one-coordinated site can either follow the stacking order dictated by its local atomic environment or it can break it, according to a user-defined probability $P\in [0,1]$ ($P=1$ ($P=0$) means that only cubic (hexagonal) stacking is allowed) \cite{LaMagna2019}. Processes where the interface kinetics is such to destabilize higher-coordinated solid sites, in favor of lower-coordinated ones, are more prone to the formation of stacking defects. 

The occupancy, coordination and bonding configuration of any bulk solid site in the super-lattice is constantly accessible in the shared-memory environment over successive FEM-KMCsL cycles. This is crucial to keep track of the amount of solid and liquid species over time, which in turn allows to compute $x_L(t)$ and $x_S({\bf r}, t)$, map the latter into the FEM solver, ensure mass conservation and track the evolution of vacancies (KMCsL voids with $n=4$ three-coordinated solid neighbours) and stacking disorder throughout the simulation.

\subsubsection*{FEM model}
Continuum modelling in this work consists in using finite element methods (FEM) to solve the heat equation self-consistently with the time-harmonic solution of Maxwell's equations on a mesh, including phase, temperature and alloy-fraction-dependent material parameters \cite{CristianoLaMagna2021}. In the context of the multiscale methodology, the mesh is 3D and the solid/liquid phase changes and species redistribution in the liquid and solid phases are modelled by means of KMCsL. In the non-atomistic simulations carried out for consistency and validation analyses, the mesh is 1D and a mixed enthalpy/phase-field formalism is adopted instead \cite{CristianoLaMagna2021, Huet2020}. The enthalpy formalism is used for $T<T_M(x)$, while the phase-field formalism is used for $T>T_M(x)$, with $T_M(x)$ being the melting temperature expected from the phase diagram. 

The mesh is always initialized from a TCAD geometry with size of $\sim$20 {\textmu}m along the direction $z$ of irradiation and 10-30 nm along the lateral $x$ and $y$ (periodic) directions. A 200 nm-thick layer of air is included above the initial surface. The KMCsL-coupled subregion typically includes the top 30-150 nm of the surface and a few-nm layer of air (2 to 10 nm), which needs to be thick enough to accommodate possible resolidification of the material above the initial surface level. The mesh resolution in the KMCsL-coupled subregion is 1-1.5 nm and gradually becomes coarser far from it, until it reaches the mesh lateral size in the top and bottom of the mesh. 

The optical and thermal parameters for Si and SiO$_2$ are taken from Ref. \cite{LaMagnaPhaseField2004}, while those for Ge are taken from Refs. \cite{Huet2020}. The behavior of SiGe alloy is nearly ideal (Raoultian) \cite{Olesinski1984}, hence most of its properties are well approximated by a linear interpolation of pure Si and Ge ones. We recently found that the dielectric function of liquid and solid SiGe requires a more careful definition to properly capture experimentally measured reflectivities \cite{DamianoArxiv}. In this work, only for the liquid SiGe dielectric function, we choose for simplicity to use a linear interpolation between pure Si and Ge weighted on $x_L$, rescaled \emph{ad-hoc} by a factor $a_i\approx 1$ to capture the reflectivity reported in \cite{DamianoArxiv}. The $a_i$ and corresponding reflectivity values are reported in Supplementary Table S4.

The models employed for relaxed and strained SiGe only differ in the initial Ge profile and in the definition of solid-phase thermal conductivity. For strained SiGe we refer to the known expression for pure silicon \cite{LaMagnaPhaseField2004}. This is justified by the small thickness of SiGe layers compared to the substrate's in our simulations. For relaxed SiGe we use the expression reported in Ref. \cite{Wagner2006}. 

\subsubsection*{Experiments}
Relaxed thick \sige{1-x}{x} samples were prepared from two 200 mm bulk Si\hkl(001) wafers (Czochralski, p-type, 1–50 $\Omega$ cm). The epitaxy process was performed by Reduced Pressure Chemical Vapor Deposition in a Centura 5200C chamber from Applied Materials. Prior to each SiGe layer epitaxy, a $H_2$ bake (1373 K, 2 min) was done to remove the native oxide. After the surface cleaning, a graded SiGe buffer layer was grown on each wafer with a 10\% / {\textmu}m ramp (1173 K for one wafer and 1123 K for the other, P = 20 Torr, precursors: SiH$_2$Cl$_2$ + GeH$_4$). Then, 1.2 {\textmu}m thick relaxed and undoped SiGe layers were grown with a uniform Ge content, corresponding to that of the buffer layer underneath. Thanks to the high temperature used during the process, the glide of the threading arms of misfit dislocations (i.e. threading dislocations) was enhanced in such way that they remained mostly confined in the graded buffer layers, close to the SiGe/Si interface. As a result, the threading dislocations density was significantly reduced in the SiGe top layers ($\sim$10$^5$ cm$^2$). Following the RPCVD process, the remaining cross-hatch patterns were removed using a two steps (planarization and smoothing) Chemical-Mechanical Polishing process thanks to a Mirra CMP system from Applied Materials, reducing the thickness of the SiGe top layers from 1.2 {\textmu}m to ($\sim$0.7 {\textmu}m). 
Nanosecond Laser Annealing was performed with a SCREEN-LASSE (LT-3100) UV laser ($\lambda$ = 308 nm, single pulse, pulse duration = 160 ns, 4 Hz repetition rate, $<3$ \% laser beam uniformity, $10 \times 10$ mm$^2$ laser beam) at room temperature and atmospheric pressure, with an constant incident N$_2$ flux to strongly limit the oxygen incorporation. The Ge composition of laser irradiated SiGe layers was measured with well-calibrated \cite{DamianoArxiv} Energy Dispersive X-ray spectroscopy (EDX) in a Transmission Electron Microscope JEM-ARM200F Cold FEG equipped with a EDX SDD CENTURIO-X detector from JEOL. X-ray signals in selected areas have been quantified via the Cliff and Lorimer factor method to extract Ge content profiles as function of depth. The cross-section lamellas were fabricated by Focused Ion Beam in a Helios 450S Scanning Tunneling Electron Microscope from FEI. 

\section*{Code availability}
The code presented in this work is fully open-source. It is available on Github and can be accessed via this link: \url{https://github.com/MulSKIPS}. 


\section*{Data availability}
The source code and minimal input datasets needed to replicate the findings reported in the article are available on Github and can be accessed via this link: \url{https://github.com/MulSKIPS/MulSKIPS/tree/main/examples/ex4-LA-SiGe}.

\section*{Acknowledgements}
The authors thank the European Union’s Horizon 2020 Research and Innovation programme under grant agreement No. 871813 MUNDFAB, and the European Union’s NextGenerationEU under grant agreement CN00000013 - NATIONAL CENTRE FOR HPC, BIG DATA AND QUANTUM COMPUTING for computational support. 

\section*{Author contributions}
G.C. and A.L.M. conceived the multiscale strategy and developed the code.
G.C. performed the multiscale simulations, prepared the figures and wrote the main text.
D.Raciti, G.C. and G.F. calibrated the atomistic part of the code.
D.Ricciarelli and I.D. calibrated the continuum part and performed the phase-field simulations. 
P.A.A., F.C., R.Daubriac, R.Demoulin, J.M.H. and S.K. provided the experimental data.
All authors discussed the results and reviewed the manuscript.

\section*{Competing interests}
The authors declare no competing interests.


\clearpage

\appendix

\renewcommand{\thetable}{\textsc{S}\arabic{table}}
\renewcommand{\thefigure}{\textsc{S}\arabic{figure}}
\renewcommand{\thesubsection}{\textsc{Note S}\arabic{subsection}}
\setcounter{figure}{0}    

\section*{SUPPLEMENTARY INFORMATION}
\subsection{Technical implementation of the multiscale method}

\begin{figure*}[t]
  \centering
  \includegraphics[width=0.98\linewidth]{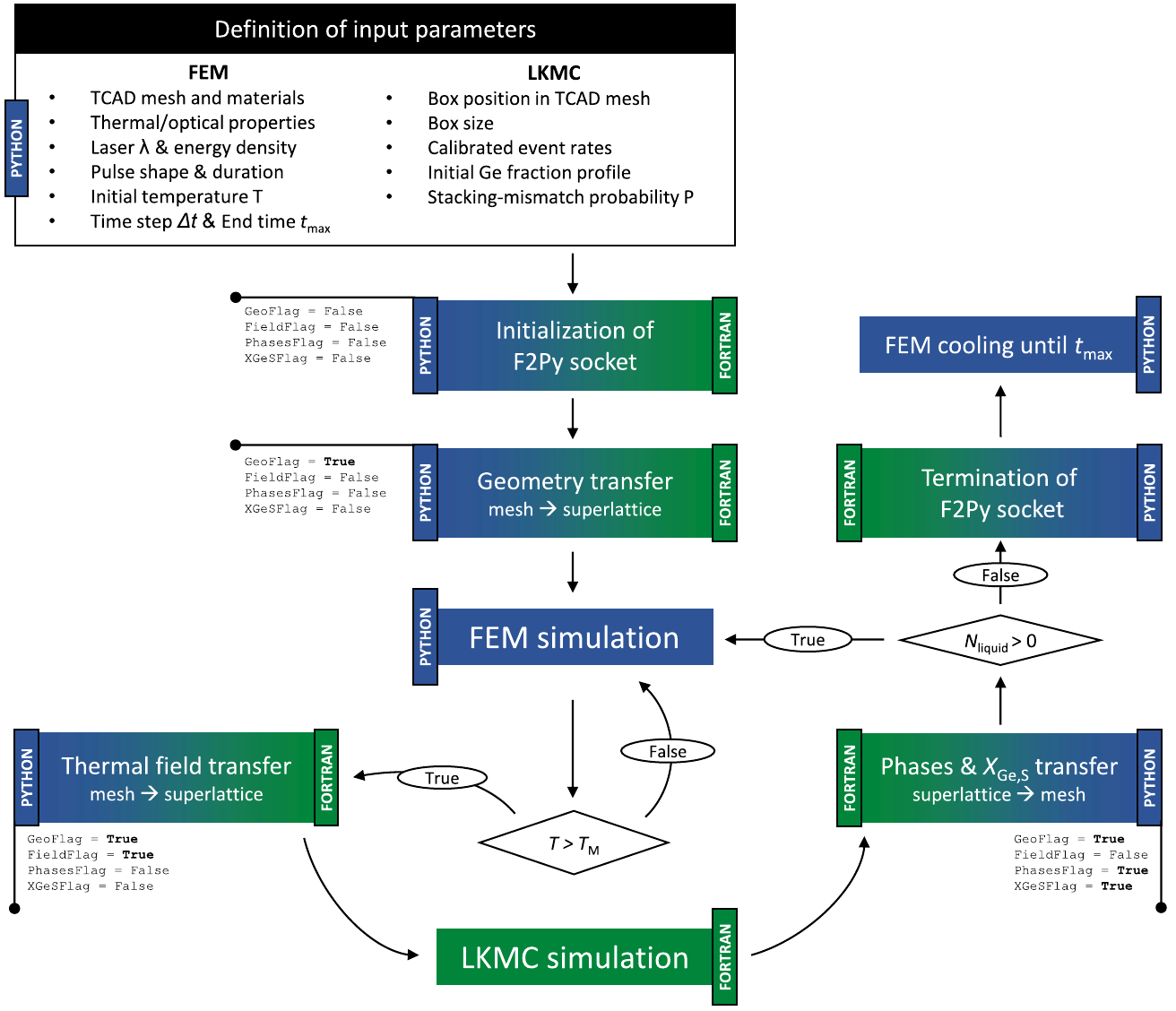}
  \caption[Technical flowchart of the hybrid atomistic-continuum tool.]{Technical flowchart of the hybrid atomistic-continuum tool. The required input parameters for FEM and KMCsL models, listed on top, are all provided to a general Python control script. Green and blue colours indicate Fortran and Python-based character of the invoked routines, respectively. Steps directly involving F2Py routines are shown with a colour gradient ranging from green to blue, and vice-versa. The control flags GeoFlag, FieldFlag, PhasesFlag and XGeSFlag are used in Fortran to signal the availability in RAM of the relative data arrays and thus trigger the communications between Fortran and Python.}
  \label{fig:figS1}
\end{figure*}

Here we report on the main software engineering tasks needed to enable nanosecond pulsed laser annealing (LA) simulations for SiGe alloys. The means and importance of sharing of RAM between Python and Fortran modules will be deployed. This gives the possibility to ensure mass conservation during the LA process, enable solidification of the material above its as-grown surface level, model extended defects during the process and track the kinetics and redistribution of the various species composing the alloy. 
The major technical improvement of enabling shared RAM, and thus efficient in-memory data transfers, between the Fortran kinetic monte carlo on super-lattice and the Python continuum FEM environments for LA simulations was achieved by integrating the so-called F2Py sockets \cite{f2py1} into the \mulskips code, exploiting some of the routines available in \cite{f2py2}. The workflow after full integration of F2Py functionalities in the Python and Fortran  modules is outlined in \Fref{fig:figS1}.
During a hybrid FEM-KMCsL LA simulation three key data transfers are carried out between the Fortran and Python solvers, one at the beginning of the simulation and the others at every time step $\Delta t$ of the FEM-KMCsL cycle. The first one is a geometry transfer from Python to Fortran, occurring after interpolation of the relevant part of mesh into the \mulskips superlattice formalism. The second one is the thermal field transfer from Python to Fortran at every time step $\Delta t$, which is used as input in \mulskips to determine the KMCsL space- and time-dependent melting/solidification event probabilities. The third one is the solid/liquid phases transfer from Fortran to Python at every time step $\Delta t$, which is used to get the phase-changed volume in the FEM model and compute the exchanged latent heat needed to solve the Maxwell-Fourier self-consistent problem, and get an updated thermal field in the following iteration. Without implementing F2Py sockets \cite{Calogero2022}, such data (in the form of arrays with KMCsL superlattice dimension $L_x\times L_y\times L_z$) could only be transferred from Python to Fortran and vice versa by means of hundreds of heavy write/read operations on disk, with obvious performance limitations and demanding storage requirements. F2Py sockets allow performing such array transfers directly in RAM.

From a more technical point of view, the operations involving F2Py sockets are performed by ad-hoc “send” and “receive” routines implemented in both Fortran and Python environments, whose execution is triggered in Fortran by proper combinations of the four control flags GeoFlag, FieldFlag, PhasesFlag and XGeSFlag, each indicating the availability in RAM of the data arrays describing occupancies, thermal field, phases and Ge local fraction, respectively. The initialization of the F2Py socket and the \mulskips unique Fortran CPU process is carried out right after setting up the LA simulation parameters in Python. At the beginning of the Fortran CPU process, all flags are set to False, meaning that the KMCsL code is on standby and waiting for instructions from Python. The Python module interpolates a local mesh region into a data array in the form of superlattice sites occupations, issues a “send” command to communicate the geometry information via the F2Py socket, and is put on standby. The “send” command in Python triggers a “receive” command in Fortran, which reads the data array stored in RAM by Python, uses them to set up the superlattice occupations, then sets the GeoFlag to True and is put in standby. In the latter step, the original number of each atomic species is also stored, as a reference to ensure mass conservation during solidification at the latest stages of the LA simulation. The Python module enters the FEM cycle and heats up the system by simulating laser absorption until the maximum temperature in the mesh reaches the trigger temperature melting temperature $T_M$. At this point, the thermal field is interpolated and sent to Fortran via F2Py sockets, which receive it and set the FieldFlag to True. This last statement, combined with the True state of GeoFlag, triggers the setup of KMCsL probabilities and marks the beginning of the first KMCsL cycle. Once the KMCsL simulated time reaches $\Delta t$, the solid/liquid state of each superlattice site and the Si/Ge identities of all solid ones are stored in RAM, a “send” command is issued (this time in Fortran), the PhasesFlag and XGeSflag are set to True and the FieldFlag is restored to False, ensuring that the Fortran CPU process stays in standby until a new thermal field is transferred. Python reads such data from RAM, uses it to update the S/L volumes in the mesh and the solid-phase local Ge fraction, then recalculates the thermal field. As this sequence goes on, the number of liquid and solid sites in the KMCsL superlattice is continuously tracked in the Fortran environment, differentiating the various chemical species (e.g., Si and Ge in SiGe alloys). Once the original number of solid species is recovered (or, equivalently, there are no more liquid sites) the backward communications to Python occur one last time and the F2Py connection is closed. At this point, the FEM cooling cycle continues until the maximum simulation time $t_{\rm max}$, set up at the beginning of the LA simulation, is reached.

In addition to optimizing data exchange, an F2Py socket allows for the execution of a unique KMCsL CPU process for the whole LA simulation (rather than a sequence of independent KMCsL calculations reinitialized at every time step $\Delta t$). This in turn means that the information about KMCsL sites’ occupations and coordinations can be retained across subsequent KMCsL-FEM communication cycles. In Ref. \cite{Calogero2022} this was unfeasible because \mulskips needed to be reinitialized after every FEM step, causing an inevitable reset of \mulskips superlattice information. 

Overall, RAM storage through F2Py sockets has the important advantages of unlocking the simulation of defects in the irradiated material, which is a native characteristic feature of \mulskips. It indeed allows preserving the information about vacancies positions in the KMCsL box across subsequent KMCsL cycles, which would otherwise be lost if \mulskips is reinitialized from scratch at every $\Delta t$. It makes it possible to keep the information about stacking choices made for all one-coordinated solid sites across subsequent KMCsL $\Delta t$-long cycles, enabling the evolution of extended stacking defects during the LA-induced resolidification.
Furthermore, it allows tracking the concentration and position of every solid species within the KMCsL box during the LA simulation, which is crucial to count solid sites at every instant of the simulation and ensure that solidification ends whenever the original solid mass is recovered.

\subsection{Details on KMCsL calibration}
The SiGe KMCsL event probabilities for species $i$ ($i=$Si,Ge) are expressed as follows.
\begin{itemize}
    \item Solidification: 
        \begin{equation}
        \nu_{\rm LS}^i = f^i(T) \cdot X^i \cdot \nu_0^i  \cdot {\rm exp}\left(-\frac{2 E_{\rm LS}^i (n)}{k_B T_M^i}\right) \label{eq:1}
        \end{equation}
        where $\nu_0^i$ and $E_{\rm LS}^i (n)$ are the Boltzmann prefactor and the energy barriers (dependent on the coordination number $n$, see \Tref{tableS1}) for the species $i$; $f^i(T) = 1/2 [1+{\rm erf}((T-T_0^i)/A^i )]$ is a damping factor; $T_M^i$ is the melting temperature of $i$. All these parameters are obtained from the calibration of pure Si and Ge melting-solidification kinetics against the respective Fulcher-Vogel curve. $X^i=1-x_L$ for Si and $X^i=x_L$ for Ge in the assumption of fast diffusion and infinite liquid reservoir.
    \item Melting: 
        \begin{equation}
            \nu_{\rm SL}^i = \nu_0^i  \cdot {\rm exp}\left(-\frac{n E_{\rm SL}^i (n_{\rm Si}, n_{\rm Ge})}{k_B T}\right) \label{eq:2}
        \end{equation}
        where $\nu_0^i$ is defined as for solidification and $E_{\rm SL}^i (n_{\rm Si},n_{\rm Ge} )$ are the melting energy barriers (equivalent to bonding energies), that depend on the number and kind of nearest neighbors ($n_{\rm Si}+n_{\rm Ge}=n$). The definition of $E_{\rm SL}^i (n_{\rm Si},n_{\rm Ge} )$ is crucial for the calibration. Under the assumption of ideal (Raoultian) mixing, which holds for the case of SiGe alloys, the energy $E_{ij}$ of a mixed bond can be estimated as $E_{ij}=((E_{ii}+E_{jj}))⁄2$. An initial guess for the values of $E_{\rm SL}^i (n_{\rm Si},n_{\rm Ge} )$ is therefore obtained as linear combinations of the solidification energy barriers for pure Si and Ge, $E_2^{\rm Si}\equiv E_{\rm LS}^{\rm Si}(2)$ and $E_2^{\rm  Ge}\equiv E_{\rm LS}^{\rm  Ge}(2)$ weighed on the number of Si and Ge nearest neighbors. The energy barriers for mixed bonding states are perturbed (i.e., further decreased in the case of Si and increased in the case of Ge) by a small amount to reproduce the experimental phase diagram of the alloy. The resulting 18 energy parameters (9 for Si and 9 for Ge) take the form reported in \Tref{tableS2}. For the sake of simplicity, the perturbation was chosen to be the same for every coordination state of Si ($\alpha$) and Ge ($\beta$), reducing the calibration to two parameters only.         
\end{itemize}

\begin{table*}[h!]
\caption{Solidification energies for Si (left column) and Ge (right column). The energy values $E_2^{\rm  Si}$ and $E_2^{\rm  Ge}$ are obtained by calibrating the pure species.}
\begin{ruledtabular}
\begin{tabular}{p{0.4\textwidth}p{0.4\textwidth}}
    Si    &   Ge      \\ 
    \hline 
    $E_{\rm LS}^{\rm Si} (1)=(E_2^{\rm Si}-\delta^{\rm Si} )$    &   $E_{\rm LS}^{\rm  Ge} (1)=(E_2^{\rm  Ge}-\delta^{\rm  Ge} )$      \\
    $E_{\rm LS}^{\rm Si} (2)=E_2^{\rm Si}$            &   $E_{\rm LS}^{\rm  Ge} (2)=E_2^{\rm  Ge}$              \\
    $E_{\rm LS}^{\rm Si} (3)=(E_2^{\rm Si}+\delta^{\rm Si} )$	&   $E_{\rm LS}^{\rm  Ge} (3)=(E_2^{\rm  Ge}+\delta^{\rm  Ge} )$
\label{tableS1}
\end{tabular}
\end{ruledtabular}
\end{table*}

\begin{table*}[h!]
\caption{Melting energies $E_{\rm SL}^i (n_{\rm Si},n_{\rm Ge} )$ for i=Si (left column) and i=Ge (right column).}
\begin{ruledtabular}
\begin{tabular}{p{0.4\textwidth}p{0.4\textwidth}}
    Si    &   Ge      \\
    \hline 
    $E_{\rm SL}^{\rm Si} (1,0)=E_2^{\rm Si}$                        &   $E_{\rm SL}^{\rm  Ge} (1,0)=(1+\beta)  [E_2^{\rm  Ge}+E_2^{\rm Si} ]⁄2$     \\
    $E_{\rm SL}^{\rm Si} (0,1)=(1-\alpha)  [E_2^{\rm Si}+E_2^{\rm  Ge} ]⁄2$     &   $E_{\rm SL}^{\rm  Ge} (0,1)=E_2^{\rm  Ge}$                        \\
    $E_{\rm SL}^{\rm Si} (1,1)=(1-\alpha)  [3E_2^{\rm Si}+E_2^{\rm  Ge} ]⁄4$    &   $E_{\rm SL}^{\rm  Ge} (1,1)=(1+\beta)  [3E_2^{\rm  Ge}+E_2^{\rm Si} ]⁄4$    \\
    $E_{\rm SL}^{\rm Si} (2,0)=E_2^{\rm Si} $                       &   $E_{\rm SL}^{\rm  Ge} (2,0)=(1+\beta)  [E_2^{\rm  Ge}+E_2^{\rm Si} ]⁄2$     \\
    $E_{\rm SL}^{\rm Si} (0,2)=(1-\alpha)  [E_2^{\rm Si}+E_2^{\rm  Ge} ]⁄2$     &   $E_{\rm SL}^{\rm  Ge} (0,2)=E_2^{\rm  Ge}$                        \\
    $E_{\rm SL}^{\rm Si} (2,1)=(1-\alpha)  [5E_2^{\rm Si}+E_2^{\rm  Ge} ]⁄6 $   &   $E_{\rm SL}^{\rm  Ge} (2,1)=(1+\beta)  [2E_2^{\rm  Ge}+E_2^{\rm Si} ]⁄3$    \\
    $E_{\rm SL}^{\rm Si} (1,2)=(1-\alpha)  [2E_2^{\rm Si}+E_2^{\rm  Ge} ]⁄3$    &   $E_{\rm SL}^{\rm  Ge} (1,2)=(1+\beta)  [5E_2^{\rm  Ge}+E_2^{\rm Si} ]⁄6$    \\
    $E_{\rm SL}^{\rm Si} (3,0)=E_2^{\rm Si}$                        &   $E_{\rm SL}^{\rm  Ge} (3,0)=(1+\beta)  [E_2^{\rm  Ge}+E_2^{\rm Si} ]⁄2$     \\
    $E_{\rm SL}^{\rm Si} (0,3)=(1-\alpha)  [E_2^{\rm Si}+E_2^{\rm  Ge} ]⁄2$     &   $E_{\rm SL}^{\rm  Ge} (0,3)=E_2^{\rm  Ge}$                        
\label{tableS2}
\end{tabular}
\end{ruledtabular}
\end{table*}

The calibration of Si-Ge mixtures is based on the single-species calibrations of pure Si and Ge. The single-species solidification-melting process simulations were calibrated against the Fulcher-Vogel relation for the pure (crystalline) species, which gives the temperature dependence of the interface velocity \cite{LaMagnaPhaseField2004, Mittiga2000}. No such relations exist in the case of alloys. On the contrary, we relied on the experimental phase diagram of the material, expressing the temperature-dependent composition of the solid and the liquid phases at equilibrium, when the melting/solidification process is operated at a very slow speed \cite{Olesinski1984}. The behavior of the alloy is very close to ideal (Raoultian), where enthalpic contributions to the mixing energy are negligible and the lens shape of the diagram mostly depends on the entropy of fusion of the component species (the larger the entropy, the broader the shape).

For calibration purposes, a few assumptions were made in the simulation set-up. The liquid phase surmounting the solid material is assumed to have a fixed Ge fraction $x_L$, meaning negligible diffusion times of Si and Ge atoms within an infinite liquid reservoir. The temperature is kept constant and uniform in the simulation box at every KMCsL run. 
We note that in the FEM-KMCsL LA simulations the KMCsL event rates in \Eref{eq:1} and \Eref{eq:2} are functions of time $t$ and lattice position ${\bf r}$. This is because the condition of fixed and uniform temperature in the simulation box is dropped and the liquid-phase composition $x_L\equiv x_L (t)$ (i.e., the parameter $X^i$) becomes a time-dependent variable. 

\begin{table*}[h!]
\caption{KMCsL calibrated parameters for pure Ge and Si.}
\begin{ruledtabular}
\begin{tabular}{p{0.3\textwidth}p{0.3\textwidth}p{0.3\textwidth}}
    - &   Ge    &   Si      \\
    \hline 
    $ν_0$             &       $2.60\cdot 10^{16}$   &       $1.33 \cdot 10^{17}$        \\
    $E_2$ [eV]        &       0.65                &       0.96                      \\            
    $\delta$ [eV]     &       0.02                &       0.03                      \\            
    $T_0$ [K]         &       900                 &       1080                      \\            
    $A$ [K]           &       210                 &       280                       \\            
    $T_M$ [K]         &       1210                &       1688                      
\label{tableS3}
\end{tabular}
\end{ruledtabular}
\end{table*}

\begin{figure*}[h!]
  \centering
  \includegraphics[width=0.95\linewidth]{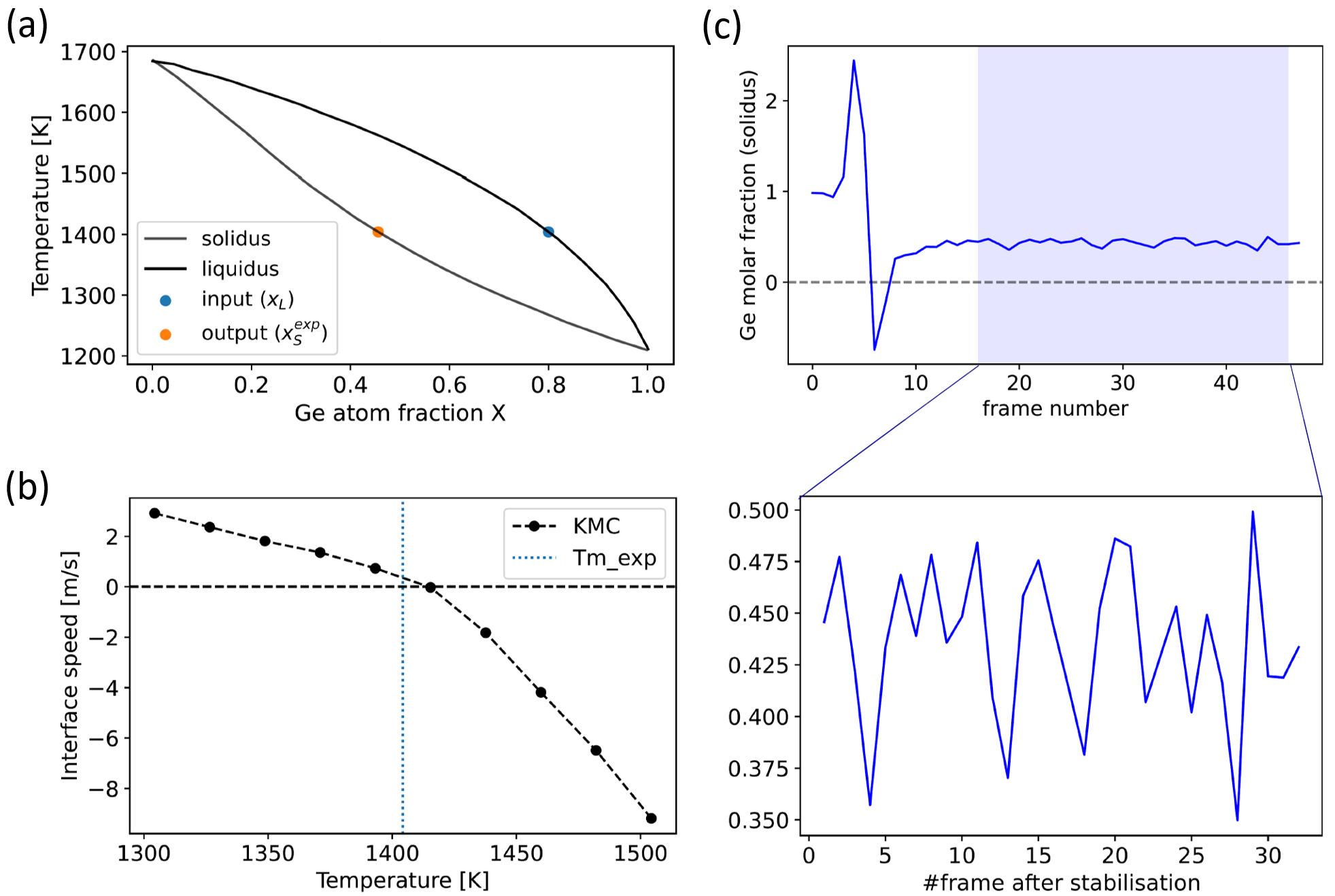}
  \caption[KMCsL calibration workflow.]{Some steps of the KMCsL calibration workflow for the case of $x_L=0.8$. (a) experimental phase diagram of the SiGe mixture. The $x_L$ value on the liquidus curve is marked in blue, the expected $x_S^{\rm exp}$ value on the solidus curve (0.46 in this case) is marked in orange. (b) temperature-dependent interface speed predicted from the KMCsL model (black dashed line). The expected equilibrium temperature $T_M^{\rm exp}$ at $x_L$ extracted from the phase diagram is shown as blue dotted line. The equilibrium temperature $T_M^{\rm KMCsL}$ interpolated from the KMCsL curve is 1415 K. (c) Evolution of the bulk Ge fraction obtained from the KMCsL run at $T_M^{\rm run}=$1410K; the initial frames are discarded until the $x_S^{\rm KMCsL}$ value stabilises around a constant value. In this case, $x_S^{\rm KMCsL}=0.44$, with standard deviation $\sigma(x_S^{\rm KMCsL})=0.04$.}
  \label{figcalib}
\end{figure*}

The overall workflow of the calibration is described below:
\begin{itemize}
    \item Set the calibration parameters $\alpha$ and $\beta$.
    \item Set the desired molar fraction of Ge in the liquid phase, $x_L$, which is assumed to be fixed during the KMCsL run and set to be equal to the initial solid seed (the latter assumption is arbitrary, since the composition of the solidified bulk material will depend on $x_L$ and on the KMCsL event probabilities). 
    \item Run \mulskips in the temperature interval $[T_M^{\rm exp}-100, T_M^{\rm exp}+100]$, where $T_M^{\rm exp}$ is the equilibrium temperature in phase diagram, corresponding to $x=x_L$ (see \Fref{figcalib}a). The equilibrium temperature, $T_M^{\rm KMCsL}$, is interpolated as the temperature corresponding to zero interface velocity in the KMCsL runs (see \Fref{figcalib}b). 
    \item Run \mulskips at a constant temperature $T^{\rm run}$ slightly lower than $T_M^{\rm KMCsL}$ (by 5 K), to ensure quasi-equilibrium solidification of the material.
    \item The Ge molar fraction of the solidified material, $x_S^{\rm KMCsL}=n_{\rm Ge}/(n_{\rm Si}+n_{\rm Ge})$, and the standard deviation $\sigma(x_S^{\rm KMCsL})$, are calculated by tracking the number of Si and Ge atoms solidified at every time frame in the KMCsL cycle, and averaging the Ge fraction once the solidification is stabilized (see \Fref{figcalib}c). 
    \item Compare the KMCsL results to the experimental phase diagram. Repeat the steps above with updated $\alpha$ and $\beta$, if necessary.
\end{itemize}

The obtained calibration parameters reproducing the Fulcher-Vogel profile for pure Si and Ge are reported in \Tref{tableS3}.
The parameters $\alpha$ and $\beta$ were set to 0.06 and 0.08 respectively, meaning a small (less than 10\%) perturbation to the Raoultian behavior of the mixture.

\clearpage
\subsection{Differences between phase-field and FEM-KMCsL approaches}

Besides the trivial difference in dimensionality and in the modelling approach to phase transitions, other technical differences exist between the phase-field and the multiscale FEM-KMCsL methodology, which are at the origin of the deviations observed in Fig. 3 of the main text. 
\begin{enumerate}
    \item Melting in phase-field simulations occurs at $T \geq T_M(x_L)$, with $T_M(x_L)$ being the liquidus line of the phase diagram drawn in Fig. 2 of the main text (dashed lines). In the FEM-KMCsL simulations, local melting is governed by the balance between the well-calibrated KMCsL melting and solidification events.

    \item The interface smoothness within the phase-field formalism is user-defined. The melt front evolution in planar samples is tracked by looking at the flex of the phase function as a function of depth. Both this phase profile and the local Ge concentration profile in the mesh are seamless and smooth step-like functions of depth, uniquely defined across solid and liquid regions. In the KMCsL formalism the solid/liquid (S/L) interface is atomically sharp, with its position in planar samples evaluated as the average $z$ coordinate of all undercoordinated solid atoms. This different smoothness is at the origin of the slightly higher (~$6$ nm) maximum melt depth found with FEM-KMCsL for 30nm strained SiGe irradiated with high energy density (i.e., 1.9 J cm$^{-2}$). Here there is an instant where the S/L interface falls below the initial SiGe/Si junction. At this point, in the phase-field simulation, the Ge concentration drops to zero in a gradual fashion. Instead, in the FEM-KMCsL simulation, a higher concentration of Ge (i.e., lower $T_M$) exists in the liquid region right above the SLI (due to its much more step-like Ge profile). The heat to be released to stop melting and initiate solidification is therefore higher in the FEM-KMCsL than in phase-field simulations.

    \item Atomic sites in the liquid phase are not explicitly implemented in the current KMCsL framework, and neither are liquid-phase diffusion events. Si and Ge concentrations are computed as averages, by direct subtraction of the number of Si and Ge solid sites from the respective initial amounts. On one hand, this averaging is justified by the high value of Ge diffusivity in liquid SiGe near melting temperature (~$10^{-4}$ cm$^2$/s \cite{LaMagnaPhaseField2004, Cai2000}), which yields Ge diffusion velocities ~3-4 times higher than the typical solidification velocity in LA processes with ns pulses (~2-3 m/s) \cite{Dagaultphdthesis, Baeri1996}. In fact, melt depths and overall Ge segregation trends are well reproduced. On the other hand, approximating as infinite such a high diffusivity leads to underestimated Ge trapping at the beginning of solidification, compared to phase-field. This is evident from the different minima found in the post-anneal Ge profiles of Fig. 3g-h of the main text. In turn, FEM-KMCsL simulations always end with the solidification of a pure Ge capping layer (x=1) with an energy density-dependent thickness, whereas phase-field simulations provide a SiGe capping layer with energy density-dependent Ge content and an almost constant thickness. The variation of capping layer thickness in FEM-KMCsL simulations explains the the energy density-dependent plateaus found in the KMCsL melt depth profiles. The variation in Ge content in the phase-field simulations explains the energy density-dependent plateaus found in the phase-field $T_{\rm max}$ profiles.  
    
    \item  The purely continuum methodology presents numerical instabilities at the onset of melting and at the end of solidification. For example, one may notice the abrupt melt depth variation in the latest stage of solidification in Fig. 3c of the main text. These instabilities are due to the switching between enthalpy and phase-field continuum solvers occurring every time $T$ crosses $T_M(x_L)$. A good convergence of the simulations needs to be achieved, by proper adjustments of the time step $\Delta t$, interface smoothness and threshold size for nucleation. The only numerical instabilities affecting the FEM-KMCsL simulations concern temperature oscillations, like the small ones observed at the onset of melting in Fig. 3e of the main text. These are due to the small (~$1.5$ nm) but finite thickness of the liquid layer initially defined at nucleation stage to stably initialize the melting phenomenon. Such oscillations can be reduced by setting a smaller $\Delta t$ \cite{Calogero2022}, as confirmed by the results for strained SiGe, where $\Delta t=0.25$ ns was used, instead of 0.5 ns. 
 
\end{enumerate}

\clearpage
\begin{table*}[h!]
    \caption{Correction factors $a_i$ used in the linearly interpolated expression for the dielectric constant of liquid SiGe, $\left[\epsilon_{l,SiGe} (x_{Ge})=\epsilon_{l,Ge} \cdot x_{Ge} + \epsilon_{l,Si} \cdot (1-x_{Ge})\right] \cdot a_i$, in order to reproduce the experimental melt depths in Figure 4 of main text. The values of reflectivity are also reported, before ($R_{a_i}=0$) and after ($R_{a_i}$) the correction. ED stands for energy density.}
    \begin{ruledtabular}
     \begin{tabular}{p{0.23\textwidth}p{0.15\textwidth}p{0.22\textwidth}p{0.15\textwidth}p{0.1\textwidth}p{0.1\textwidth}}
        Sample     &     $\Delta t_{\rm pulse}$ [ns]    &    ED [J cm$^{-2\,}$]  &    $a_i$  &    $R_{a_i}$   &    $R_{a_i=0}$  \\[2pt] 
        \hline 
        Relaxed \sige{0.24}   &   160   &   0.75    &   1.5627   &   0.815    &   0.778    \\
                              &   160   &   1.10    &   1.4160   &   0.807    &   0.778    \\
        Relaxed \sige{0.58}   &   160   &   0.90    &   1.3258   &   0.802    &   0.775    \\
        Strained \sige{0.2}   &   146   &   1.80    &   1.1938   &   0.794    &   0.778    \\
                              &   146   &   2.00    &   1.4073   &   0.807    &   0.778    \\
        Strained \sige{0.4}   &   146   &   1.60    &   1.4073   &   0.769    &   0.768    \\
                              &   146   &   1.81    &   1.2460   &   0.797    &   0.768    
    \label{table:1}
    \end{tabular}
   \end{ruledtabular}
\end{table*}

\begin{figure*}[h!]
  \centering
  \includegraphics[width=0.95\linewidth]{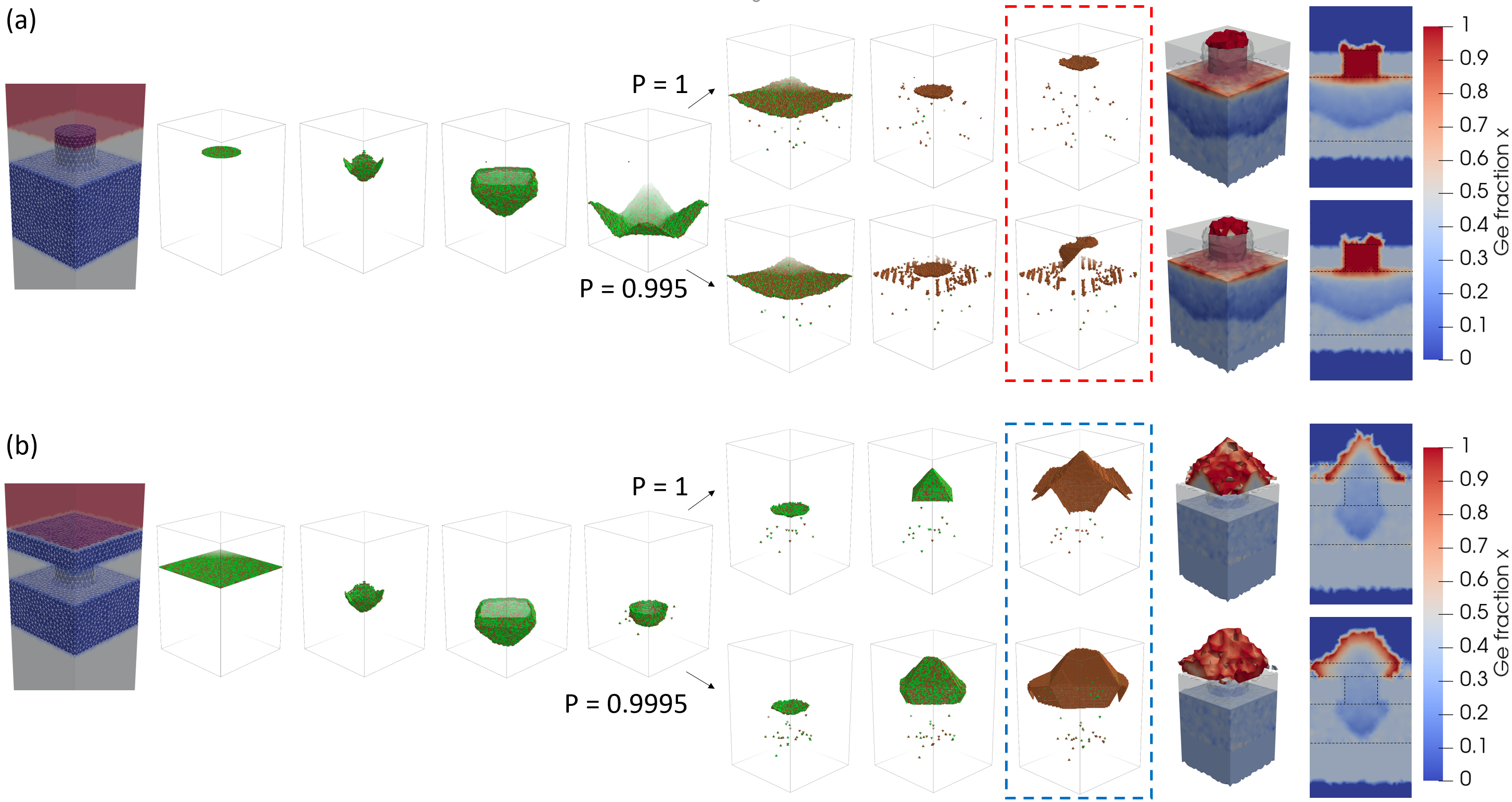}
  \includegraphics[width=0.95\linewidth]{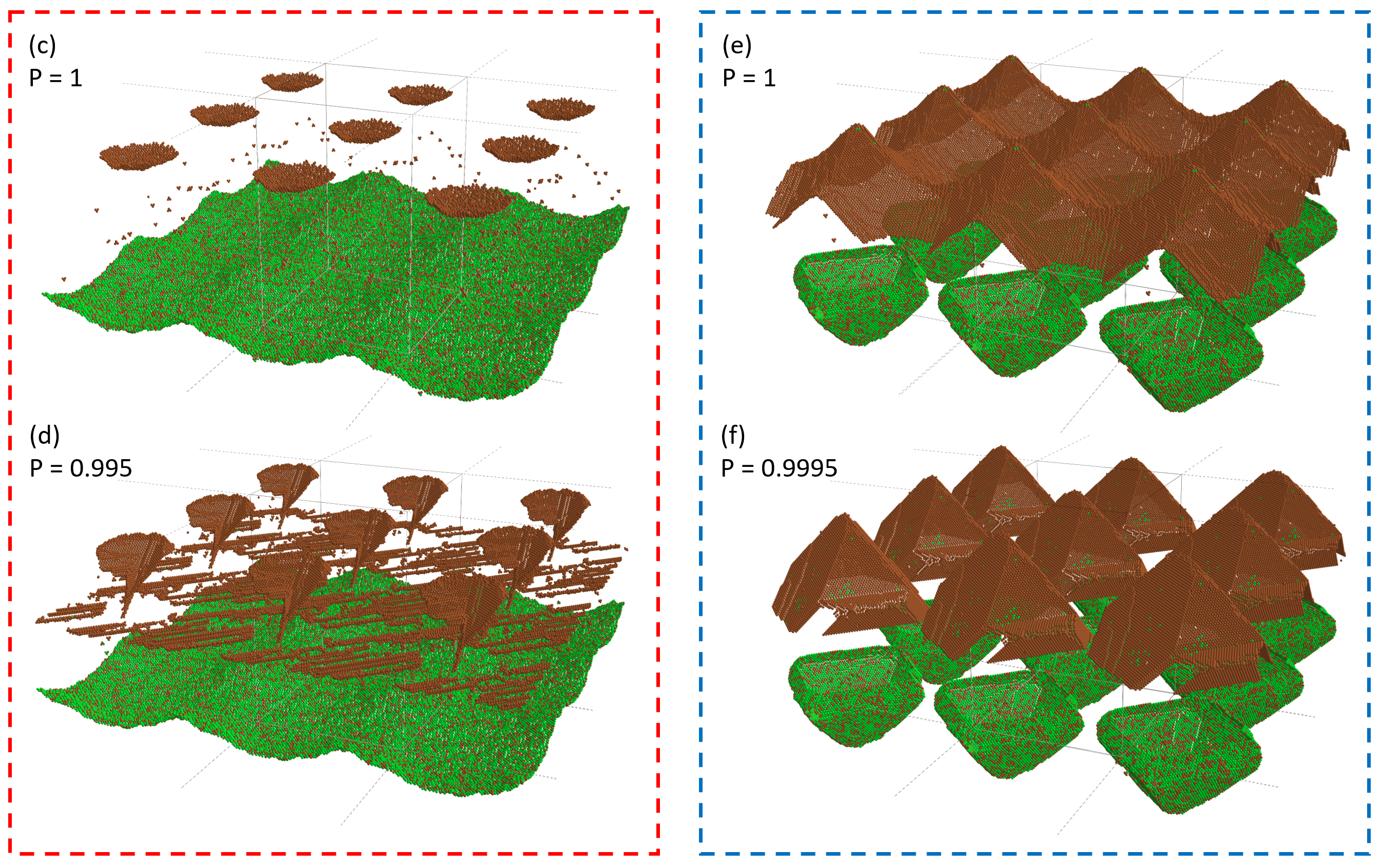}
  \caption[LA simulation results for advanced SiGe systems]{LA simulation results for advanced SiGe systems. (a-b) Same simulations of Figure 6 in main text, but with KMCsL snapshots not overlapped. A $3\times3$ periodic repetition of the KMCsL interface at the end of melting and solidification stages in the simulations in (a) is shown in (c) for P=1 and in (d) for P=0.995. For the simulations in (b), the KMCsL cell is reported in (e) for P=1 and in (f) for P=0.9995.
  }
  \label{fig:fig3x3}
\end{figure*}

\end{document}